\documentclass[12pt,preprint]{aastex}

\usepackage{lscape}

\def\farcs{\hbox{$.\!\!^{\prime\prime}$}}

\def\half{{\leavevmode\kern.1em\raise.5ex
\hbox{\the\scriptfont0 1}\kern-.1em /
\kern-.15em\lower.25ex\hbox{\the\scriptfont02}}} %exercise 11.6
\def\gtsim{\lower.5ex\hbox{$\buildrel > \over\sim$}}
\def\ltsim{\lower.5ex\hbox{$\buildrel < \over\sim$}}

\def\arcs{\hbox{$^{\prime\prime}$}}

\slugcomment{To appear in The Astrophysical Journal Supplement Series}

\shorttitle{The IRAS$\,$20126 X-ray Cluster}
\shortauthors{Montes et al.}

\begin{document}

\title{X-Ray and Radio Observations of the Massive Star Forming Region IRAS$\,$20126+4104 }

\author{V. A. Montes$^{1}$, P. Hofner$^{1,\dag}$, C. Anderson$^{1,2}$, V. Rosero$^{1,2}$}

\affil{$^{1}$ Physics Department, New Mexico Tech, 801 Leroy Pl., Socorro, NM 87801, USA}
\affil{$^{2}$ National Radio Astronomy Observatory, 1003 Lopezville Road, Socorro, NM 87801, USA}
\altaffiltext{\dag}{Adjunct Astronomer at the National Radio Astronomy Observatory}

\begin{abstract}
We present results of  {\it Chandra} ACIS-I and {\it Karl G. Jansky Very Large Array} (VLA) $6\,$cm continuum observations of the IRAS$\,$20126+4104 massive star forming region. We detect 150 X-ray sources within the $17\arcmin \times 17\arcmin$ ACIS-I field, and a total of 13 radio sources within the $9\farcm2$ primary beam at $4.9\,$GHz. 
Among these are the first $6\,$cm detections of the central sources reported by Hofner et al. (2007), namely I20N1, I20S, and I20var. A new variable radio sources is also reported. Searching the 2MASS archive we identified $88$ NIR counterparts to the X-ray sources. Only 4 of the X-ray sources had $6\,$cm counterparts.
Based on an NIR color-color analysis, and on the Besan\c{c}on simulation of Galactic stellar populations (Robin et al. 2003), we estimate that about 90 X-ray sources are associated with this massive star forming region.
We detect an increasing surface density of X-ray sources toward the massive protostar and infer the presence of a cluster of at least 46 YSOs
within a distance of $1.2\,$pc from the massive protostar.

\end{abstract}

\keywords{open clusters and associations: individual (IRAS$\,$20126+4104) $-$ radio continuum: stars $-$  stars: formation $-$  stars: pre-main-sequence $-$  X-rays: stars}
\bigskip
\newpage

\section{Introduction}

The environment in which massive stars form is one of the most important parameters to decide between current models of high mass star formation. In the monolithic collapse model a massive star forms from a single massive core which mass will determine the final stellar mass (e.g. McKee \& Tan 2003). In contrast, the competitive accretion model (e.g. Bonnell \& Bate 2006) relies on enhanced mass accretion in the center region of a stellar cluster to form massive stars. In this case, the mass of the massive star is related to the properties of the cluster. It is thus interesting to study the stellar environment of massive stars, and since dynamical processes will substantially affect the appearance of a cluster, studying early phases of massive star formation is required. In fact, the tendency for massive stars to form in dense clusters is a well established observational fact. Lada \& Lada (2003), and Testi et al. (1999) reported a smooth transition in cluster richness from Ae to Be type stars, with only the most massive ($10 - 20\,$M$_\odot$) stars found in dense stellar clusters.
In contrary to these findings, several authors have also reported massive stars which apparently were formed in isolation, or in a very poor cluster environment (e.g. de Wit et al. 2005, Oskinova et al. 2013). A notable example of a such a result is the study of Qiu et al. (2008) for the case of the prominent massive protostar IRAS$\,20126+4104$,  where no obvious cluster was detected using {\it Spitzer} IRAC and MIPS observations.

Due to the large extinction in regions of massive star formation, most studies of the cluster environment have been made at infrared wavelengths (e.g. Lada \& Lada 2003). The availability of the {\it Spitzer Space Telescope}  has added much more capability to such studies, however at mid-IR wavelengths extended emission from hot dust, as well as saturation in the IRAC and MIPS instruments often cause difficulties. An alternative, and in fact complementary, approach is to make use of the high resolution imaging capacities of the {\it Chandra X-Ray Observatory}. At energies above a few keV the extinction of X-rays is low (e.g. Morrison \& McCammon 1983), and the contamination by non-cluster sources is relatively small (e.g. Getman et al. 2006). It is well known that X-ray emission from young stars is highly elevated (Feigelson \& Montmerle 1999), thus X-rays are an ideal probe of young stellar clusters. Moreover, the identification of young stars is not biased to stars with circumstellar disks as is usually the case for NIR studies, where young stars are identified by their infrared excess.

In this paper we report X-ray and radio observations toward the prominent and well studied massive protostar IRAS$\,20126+4104$, located at a distance of $1.7\,$kpc. As mentioned above, Qiu et al. (2008) found no signs for the presence of a cluster, thus raising the important question about the stellar environment in which this massive star is forming. IRAS$\,20126+4104$ is a prime example of a massive star which is currently forming through disk accretion (e.g. Cesaroni et al. 2014). The object has a luminosity of about $10^4\,$L$_\odot$, and is associated with a bipolar outflow observed in a number of molecular tracers (e.g. Cesaroni et al. 1997, Lebron et al. 2006).
At smaller scales an ionized jet is seen (Hofner et al. 1999, 2007) and a disk like structure elongated perpendicular to the flow has been detected and recently imaged with high quality in the CH$_3$CN($12-11$) transition (Cesaroni et al. 2014). Fitting Keplerian rotation curves to the molecular line data, a mass for the central object between $7 - 10\,$M$_\odot$ was derived (Cesaroni et al. 2005, Keto \& Zhang 2010).

In section~2 of this paper, we describe the observations and data reduction methods for both the {\it Chandra} and VLA observations.
Section~3 presents the observational results for X-ray and radio data, as well as archival near-infrared, mid-infrared and optical data. An analysis of  the X-ray selected near-infrared data is given in section~4. We will discuss the implications of our results in section~5, and conclude with a brief summary  in section~6.

\section{Observations and Data Reduction}

\subsection{Chandra ACIS-I}
\label{sec:Chandra ACIS-I}

The IRAS$\,$20126+4104 region of massive star formation was observed with the Advanced CCD Imaging Spectrometer (ACIS) on board the
{\it Chandra} X-Ray Observatory on March 17, 2003. The energy range of ACIS is 0.1 to 10\,keV and the total exposure time was
$39.35\,$ks.     For details on the instrument see  Weisskopf, O'Dell \& van Speybroeck (1996), Weisskopf et al. (2002), and Garmire et al. (2003).   The nominal pointing position for the ACIS array was R.A.(J2000) = $20^h14^m30\fs27$, Dec.(J2000) = $+41\degr13\arcmin42\farcs1$. The observations were taken in the standard ``Timed Event, Very Faint'' telemetry mode. The roll angle of the space craft during the observations was $58.19^\circ$, and the focal plane temperature was $-119.6\,^\circ$C. Although 6 CCD chips (I0-I3, S2, S3) were active during the observations, no useful data were obtained from the spectroscopic array and we report here data only from the imaging array ACIS-I.   The imaging array consists of four abutted $1024\times 1024$ pixel CCDs (pixel size $0\farcs492$) covering an angular region of about $17\arcmin \times 17\arcmin$.
Data reduction was performed using the CIAO software package version 3.3.01 provided by the {\it Chandra} X-ray Center, starting from level~2
reprocessed data (processing version DS 7.6.8). This version of the data processing pipeline provides an improved aspect solution and
correction of effects due to the increase of the charge transfer inefficiency (Townsley et al. 2000). We selected ASCA grades 0, 2, 3, 4, 6, and 
the data were gain-corrected and filtered for bad CCD pixels and times of bad aspect, and the energy range was restricted to $0.5 - 8\,$keV where the point-spread-function (PSF) was of good quality. Exposure maps were created and applied to the data in the standard fashion. No background flares were detected during the observations and the average background emission, as measured in a source-free region in the ACIS-I chips,
was $2.3\times 10^{-7}\,$count$\,$s$^{-1}\,$pixel$^{-1}$.

To check the astrometric accuracy of the Chandra data, the positions of $17$ bright X-ray sources located near the center of the ACIS array, were compared with their counterparts in the Two Micron All Sky Survey (2MASS, Skrutskie et al. 2006)\footnote{http://www.ipac.caltech.edu/2mass/} 
catalogue. Although the initial astrometry was already very good, with a maximum individual deviation of $< 0\farcs4$ in either coordinate, a small systematic offset of $0\farcs2$ toward the west, and $0\farcs1$ toward the south was detected. After correction of the Chandra source positions to bring it into the 2MASS/Hipparcos reference frame, the root-mean-square (RMS) offset in R.A. and Dec. between the 17 {\it Chandra} and associated central 2MASS sources, was about $0\farcs1$.

\subsection{VLA 6$\,$cm Continuum}

C-band ($6\,$cm) continuum observations of the massive star-forming region IRAS$\,$20126+4104 were conducted with the VLA operated by NRAO\footnote{The National Radio Astronomy Observatory is a facility of the National Science Foundation operated under cooperative agreement by Associated Universities, Inc.} on 2011, August 7. The phase-center of the array was located within $1\arcmin$ from the Chandra pointing position at 
RA(J2000)=$20^{h}14^{m}26^{s}.00$, Dec(J2000)=$41\degr13\arcmin33\farcs.0 $. The observations were made in the A-configuration, with the correlator covering two 1 GHz wide bands centered at 4.9 and 7.4 GHz. The FWHM of the primary beams at these frequency bands were
$9\farcm2$ and $6\farcm1$, respectively. Each band was divided into 8$\times$128 MHz spectral windows (SPWs). Therefore, the data were recorded in 16 unique SPWs, each of these with 64 channels (spectral resolution $=2\,$MHz), i.e,  a total bandwidth of 2048 MHz. Our frequency setup was chosen to avoid known sources of RFI and the strong CH$_{3}$OH $6.7\,$GHz maser. A total of 53 scans were obtained  consisting of the flux calibrator (3C286), the phase calibrator (J2007+4029) and the target source. Alternating observations between the source and the phase calibrator were made with a cycle time of $13\,$minutes and a total on-source time of $36\,$minutes.

The data were processed  using NRAO's Common Astronomy Processing System (CASA\footnote{http://casa.nrao.edu}). After flagging bad data due to band roll-off and RFI, a bandpass solution was made using 3C286. This solution was then applied when solving for the amplitude and phase corrections. The absolute flux for 3C286 was set using the Perley-Butler 2010 flux calibration standards. The complex gains derived from observations of the phase calibrator were  applied to the target source IRAS$\,$20126+4104. Images of the calibrated UV data were made using Briggs ${\tt ROBUST}=0.5$ weighting which gave a good balance between angular resolution and sensitivity. \\
Line contamination was checked by imaging each SPW separately. We made separate, primary beam corrected images of the data sets centered at 4.9 and 7.4 GHz, covering the entire FWHM of the primary beam of each band. In addition, to maximize the S/N ratio, we made an image from the combination of these two  bands. The synthesized beam of the combined image is $0.34\arcsec \times 0.29\arcsec$, sposition angle PA $=64.3^{\circ}$, and the rms noise is $6.5\mu$Jy beam$^{-1}$. \\

\section{Results}
\subsection{X-ray Sources}
\subsubsection{Detection}

Sources were identified in the ACIS-I field-of-view using WAVDETECT, a wavelet-based source detection program that works well to detect closely spaced sources (Freeman et al. 2002).  We used a  "threshold significance'' of $10^{-6}$, and wavelet scale sizes from 1 to 16 pixels incremented by a factor $\sqrt{2}$. These values provided good sensitivity to faint sources (e.g. $< 100$\,counts). To more reliably identify weak sources that only emitted in the  soft ($0.5-2\,$keV) or hard ($2-8\,$keV) X-ray energy ranges, WAVDETECT was run for each range separately as well as for the full energy range of ($0.5-8\,$keV). Spurious sources were identified visually (e.g. linear stripes or edge effects) and removed manually. Sources 
with at least 7 counts within the source detection region identified by WAVDETECT were considered to be real sources. For a thermal spectrum with energy $1.0\,$keV and extinction corresponding to a hydrogen column density (N$_H$) between $0.2 - 2\times 10^{22}\,$cm$^{-2}$, this corresponds to detection limits for the total  (i.e. absorption corrected) luminosity of $5 \times 10^{29} - 3\times 10^{30}$ erg\,s$^{-1}$.

The total number of sources detected was 150. This included sources that were only detected in either the soft or hard X-ray energy ranges, as well as those that were detected in the full energy band. Approximately $80\,\%$ of the sources had fewer than $50$ counts (count rate $1.42$\, cts/ks).
The observed X-ray properties of all sources are given in Table~1. In column~1 we list the source number, column~2 is their {\it Chandra} designation according to the source position, column~3 and~4 give R.A. and Dec to higher precision, and column~5 is the observed count rate.
In column~6 we give the observed X-ray flux, column~7 is the hardness ratio and column~8 lists the variability classification. The hardness ratio (HR)  is defined as $\frac{h_{x}-s_{x}}{h_{x}+s_{x}}$ where $h_{x}$ is the count-rate in the $2-8$\,keV energy range, and $s_{x}$ is the count rate between $0.5-2$\,keV. Values of $\pm 1.00$ in the HR column thus indicate that the source was only detected in the hard or soft energy band.

Figure~1 shows a grey scale representation of the ACIS detector. The brightest detected X-ray source, nr. 148 (CXO$\,$J201514.4+411531.9),
had a total of 340 counts  and was identified with a foreground main sequence K05 spectral type star. Two prominent sources located in the core region are nr. 42 (CXO$\,$J201426.0+411331.7, 167 counts), and  source nr. 43 (CXO$\,$J201426.2+411327.9, 33 counts). The latter two sources correspond to the radio sources I20S and I20var (Hofner et al. 2007), respectively, and have recently been discussed by Anderson, et al. (2011).

\subsubsection{Timing Analysis}

Timing analysis was performed for all of the sources in the ACIS field-of-view. The count rate versus time (i.e. X-ray light-curve) was determined by measuring counts in $2000$\,s temporal bins within angular regions defined by WAVDETECT.  The background from a nearby, source-free region was then subtracted to obtain the final light-curve. Analysis of the light-curves was done using the LCSTATS program in the XRONOS package. The $\chi^{2}$ method was used to determine source variability. In the list of 150 detected sources, 13 sources exhibited variability.
Among these, four sources (11, 14, 38, 116) show flare like variability with a fast rise and a slow decay to pre-flare levels. We show these sources in Figure~2.
In Figure~3 we display the light-curves of the remaining variable sources. For these sources the variability has a different characteristics than the flare sources shown in the previous figure. The longer time scale suggests that the X-ray variability might be related to stellar rotation, or orbital modulation in a binary. The low detection rate of variability in our sample is most likely strongly affected by the low count rates of our data.

\subsubsection{Spectroscopy}

Model fitting of the spectra was performed using the XSPEC software package, using CIAO version 3.4 spectral analysis method for point-like sources. Sources were divided into three groups. First, for the majority of sources the total number of counts was not high enough that a unique model could be fitted. In that case, we used a simple thermal model with a single absorption component. The value for the temperature (T) was frozen to an arbitrary value, and the spectrum was fitted varying N$_H$ only. In this way we achieved good fits, however the physical values of the model are not meaningful, and we only report the observed fluxes in Table~1. Examples of spectra of these sources are shown in Figure~4, top panel. 
 
Second, we were able to fit one temperature (1T) thermal models with both T and  N$_H$ as free parameters to 10 sources. In Figure~4, middle panel, we show examples of these fits. The resulting fit parameters are listed in Table~2. In this table we also list the corrected (i.e. unabsorbed) values for the X-ray flux (F$_{x,c}$) and luminosity (L$_{x,c}$) in the $0.5 - 8\,$keV energy range.

Third, a small number of sources required more complicated models. In Figure~4, lower panel, we show examples of  two temperature (2T) models where we allowed both absorption components and temperatures to vary. The result of the fits are presented in Table~3.

A non-thermal (power-law) model was applied if a thermal model would not converge, and finally, a mixed model was used for sources which would not converge for either a thermal or non-thermal model alone. However, these fits were generally poorly constrained and we only list the observed fluxes for these sources in Table~1. Finally, no reliable fits were found for a total of 12 sources.

\subsection{VLA Sources}

 Source detection was made by visually inspecting the combined map of the two $1\,$GHz bands within the $6\farcm1$ FWHM of the primary beam at the higher frequency ($7.4\,$GHz). We adopted a detection level of 5 times the local rms noise level. We found 7 sources within this region, and we report the flux value as measured in the combined frequency map. Among the detected radio sources are 3 of the sources which were previously discussed by Hofner et al. (2007), namely the two ionized jet sources N1 and S, and the variable radio and X-ray source I20var which radio emission is best explained by gyrosynchrotron radiation from a low mass YSO. Our $6\,$cm observations are the first detections of these sources at this wavelength. In particular, prior to our observations the southern jet (S) had only been detected at $3.6\,$cm, and our measured $6\,$cm flux confirms that the emission is of thermal nature as was assumed by Hofner et al. (2007). We also note the discovery of a new, and relatively bright $6\,$cm source, G$78.1907+3.634$, located about $4\arcsec$ NW of the radio jet sources. With a flux density of nearly $0.2\,$mJy at $6\,$cm, for any reasonable spectral index, the source should have been detected in the deep $3.6\,$cm observations of Hofner et al. (2007). Thus G$78.1907+3.634$ is a new candidate for a variable compact radio source, which nature is possibly similar to I20var.

We also detected 3 sources outside the primary beam of the $7.4\,$GHz map, which were however located within the larger primary beam of the lower frequency band ($4.9\,$GHz). Due to their location the flux contribution from the high frequency band is uncertain, and we only report the flux measured in the low frequency band for these sources. Furthermore, we detected 3 additional  sources which were located outside the $9\farcm2$ FWHM of the $4.9\,$GHz primary beam. For these sources we report the integrated flux from this map, however due to their location outside the primary beam FWHM the reported fluxes are not very accurate. The peak positions of all sources and the integrated fluxes are listed in Table~4. With the exception of two of the sources detected outside the FWHM of the $4.9\,$GHz primary beam, the radio sources are point-like or marginally resolved. Contour maps of the resolved sources are shown in Figure~5.

\subsection{Multi-wavelength Counterparts}

In our Chandra/ACIS observations we have detected $150$ X-ray sources.  In this section we report on a search for counterparts of the X-ray sources at radio, NIR, Mid-IR, and at visible wavelengths. The identification of the X-ray source counterparts was done using the USNO-B1 catalog for visible, 2MASS catalog for NIR, and {\it Spitzer} archive for mid-IR wavelengths. The astrometric registration between these catalogs is better than $1\arcsec$, and the main limiting factor for the identification is the low number of X-ray counts for most of the {\it Chandra}/ACIS sources.

In Figure~6 we show a three-color 2MASS image of the IRAS$\,$20126+4104 region, on which we have indicated the position of the ACIS array.
We used the 2MASS All-Sky Point Source Catalog \footnote{http://irsa.ipac.caltech.edu/} to find the NIR counterparts. A matching radius of $1\arcs$ was chosen for X-ray sources with off-axis position $\theta \leq 3 \arcmin$ and was enlarged to $2 \arcs$ for $\theta > 3 \arcmin$, because of the off-axis Chandra PSF degradation (Getman et al. 2005a). With these criteria we found a total of $88$ counterparts for the $150$ X-ray sources (59$\%$). Most (90$\,\%$) of the NIR counterparts have high quality photometry in the JHK$_{s}$ bands. We list the sources and magnitudes in Table~5.

For the Mid-IR wavelength regions we searched the  {\it Spitzer} Enhanced Imaging Products Point Source catalog\footnotemark[5]. Using the same procedure as for the 2MASS catalog we found a total of $19$ counterparts for the X-rays sources within the entire ACIS field. These sources are listed in column~4 of Table~5. Qiu et al. (2008) have reported {\it Spitzer} IRAC and MIPS observations of a $5\arcmin \times 5\arcmin$ area of the IRAS$\,$20126+4104 core. We find that only 3 of their sources have X-ray counterparts and report them in Table~5. 

The optical counterparts of the ACIS sources were found using the USNO-B1.0 catalog\footnote{http://vizier.u-strasbg.fr/viz-bin/VizieR-3?-source=I/284/out}. Applying the same selection criteria as for 2MASS and {\it Spitzer}, we found that $80$ of the ACIS sources ($53\%$) have optical counterparts. These are listed in column~6 of Table~5.

We compared the VLA sources with the ACIS sources in order to find the radio counterparts of the X-ray sources. We found that 4 of the ACIS sources have radio counterparts. All the radio counterparts have position differences smaller than $1\farcs0$. The radio counterparts are also listed in Table~5

\subsection{ACIS Sources Without 2MASS Counterparts}

Extragalactic contamination at X-ray wavelengths in the galactic plane due to AGNs was simulated by Getman et al. (2006) and Wang et al. (2006) at nearly the same galactic latitude as IRAS$\,$20126+4104 and at a similar exposure time. They predict that 20 $\pm$ 10 sources detected should be extragalactic sources, with a only few having NIR counterparts. In our Chandra observations, we obtain 62 X-ray sources without 2MASS association. Comparing the properties of X-ray sources with and without 2MASS counterparts, we found that the X-ray sources without 2MASS counterparts are more uniformly distributed across the ACIS field than the X-ray sources with 2MASS counterparts which appear more clustered around the dense 
molecular core in IRAS$\,$20126+4104. This is consistent with background contamination by AGNs. In general the X-ray sources without 2MASS identifications have lower count rates (median 0.425 cts/ks) than the sources with 2MASS counterparts (median 0.675 cts/ks ). Hardness ratios (or limits) could only be determined for about $50\%$ of the sources without 2MASS counterparts, the data indicating about equal numbers of hard and soft sources.  

Thus, a reasonable explanation for sources without NIR counterparts is that they consist to approximately equal parts of AGNs and late type stars which would be below the sensitivity limit of 2MASS at the distance to IRAS$\,$20126+4104. An additional, but relatively small component of non-NIR detected X-ray sources are those near the massive protostar which are not detected due to the reflection nebulosity in the center.

\section{Analysis}

\subsection{NIR Properties of the X-Ray Sources}

Inspecting the stellar field toward IRAS$\,$20126+4104 (Figure~6) the presence of a cluster is not obvious in the crowded environment of the Cygnus region. However, at the distance of IRAS$\,$20126+4104 ($1.7\,$kpc) most of the detected X-ray sources will be young stars, and we can obtain a view of the stellar content surrounding the massive protostar in the region by selecting the X-ray detected 2MASS sources.

In Figure~7 we show the H-K$_{s}$ vs. J-H color-color diagram of all X-ray sources with 2MASS counterparts which have a high quality photometry (Table~5). The solid and dashed black lines represent intrinsic colors for main-sequence and giant stars from the compilation of Cox (2000, Tables 7.6, 7.7 and 15.7). The gray dashed lines show the reddening band for main sequence stars from Rieke $\&$ Lebofsky (1985). The dash-dotted line is the locus for classical T-Tauri stars (CTTS) from Meyer et al. (1997), and the dotted line represents the reddening band corresponding to the CTTS colors.  

Inspecting Figure~7, we see that the majority of the 79 stars are located in the color space between the two gray dashed lines, i.e. these stars can be explained by normal reddening, without any infrared excess due to a circumstellar disk. Thus the most likely explanations for these stars are weak-line T-Tauri stars (WTTS). For stars located to the right of the reddening band infrared excess is present. We find 3 sources which colors are consistent with reddened CTTS (source nr. 86, 58, 55). Sources 55 and 58 are located within $30\arcsec$ from the central protostar, and source 86 is located at a distance of $3\farcm6$.  Although these sources are weak X-ray emitters which do not allow to determine a precise value for N$_H$,
from their emission characteristics it appears that they are located outside the main molecular core where a massive star is currently forming.
We also detect one object, source 41, with a large infrared excess of H-K$_s = 2.4$. This source is located within $6\arcsec$ from the central protostar and its X-ray emission is quite weak (ctr = $0.49\,$cts$\,$ks$^{-1}$). This source is likely a class I protostar.

Figure~8 shows the J vs. J-H color-magnitude diagram for the same stars shown in Figure~7. This plot allows to constrain individual masses and reddening. The dashed line shows the location of main-sequence stars at the distance to IRAS$\,$20126+4104, and the solid black line indicates the $1\,$Myr isochrone from Siess et al. (2000). Reddening vectors with A$_V = 10$ for different stellar masses are shown as dotted lines. No foreground absorption was applied for the main-sequence and $1\,$Myr isochrone location.The majority of the stars are located to the right of the $1\,$Myr isochrone and hence trace the population of young stars in IRAS$\,$20126+4104. They appear to have masses between 0.5 M$_\odot$ $<$ M $<$ 2 M$_\odot$ reddened by 1 mag $\lesssim$ A$_V$ $\lesssim$ 5 mag. A small number of sources are also located to the left of the $1\,$Myr isochrone. We have used the Besan\c{c}on simulation of Galactic stellar populations (Robin et al. 2003) to determine the location of expected foreground and background stars in the J vs. J-H color-magnitude diagram, and found that the sources to the left of the $1\,$Myr isochrone, approximatively 30 sources, are most likely due to foreground and background contamination. In the color-color diagram (Figure~7), these stars are located near the main-sequence branch with 0.1 $ < $ J-H $ < $ 0.6 and 0.0 $ <$ H-K$_{s}  < $ 0.3.

\section{Discussion}

In this paper we have collected multi-wavelength data to investigate the stellar environment of the massive protostar IRAS$\,20126+4104$. 
This object has accumulated a mass between $7 - 10\,$M$_\odot$ and is very likely in a rapid accretion phase with an estimated accretion rate of
about $2\times 10^{-3}\,$M$_\odot \,$yr$^{-1}$ (Cesaroni et al. 2005). The object is embedded in a molecular core of approximate mass $200\,$M$_\odot$ (e.g. Shinnaga et al. 2008), and hence sufficient matter exists to further grow this protostar to yet higher mass. As discussed previously it is thus interesting to investigate the cluster environment of this massive protostar. 

Qiu et al. (2008) observed the  IRAS$\,20126+4104$ region with the {\it Spitzer Space Telescope}, and based on a color-color analysis they identified
19 young stellar objects (YSOs) within an area of $5\arcmin \times 5\arcmin$. Qiu et al. (2008) concluded from that result that IRAS$\,20126+4104$ shows no obvious cluster associated with the massive protostar. In the same area observed by Qiu et al. (2008) we identify 26 X-rays sources, however only 3 X-ray sources are coincident with the Qiu et al. (2008) sources. This result can be explained with the relatively low sensitivity of our X-ray data which only detects fairly bright X-ray YSOs, and also by the fact that YSOs without disks will in general not be identified by the Qiu et al. observations. We also find 7 compact radio sources in this area. These are likely similar objects as I20Var (see Hofner et al. 2007), namely young stellar objects emitting gyro-synchrotron radiation. Of these 7, 4 have neither {\it Chandra} nor {\it Spitzer} counterparts. Thus, combining the results of our study with that of Qiu et al. (2008) we find 46 YSOs within $1.2\,$pc from the central object. As discussed, this number is very likely a lower limit.  Comparing this result with the compilation of clusters in Lada \& Lada (2003, their Table~1), we note that this number is consistent with typical embedded clusters about B-type stars.

Another hint for the presence of a young cluster associated with the B-type protostar IRAS$\,20126+4104$ is the general distribution of X-ray sources in the region. As seen in Fig.~1, 150 X-rays sources are present over the entire ACIS field, which at a distance of $1.7\,$kpc corresponds to an $8.4\,$pc$\, \times \,8.4\,$pc area. We estimate that about 60 of these X-ray sources are either AGNs, or unrelated background/foreground objects. Hence, about 90 X-ray sources are associated with this star forming regions. This corresponds to an X-ray source surface density of $1.3\,$pc$^{-2}$. Moving to smaller scales near the massive proto-star, we find 26 X-ray sources within the region surveyed by Qiu et al. (2008), corresponding to an X-ray source surface density of $4.3\,$pc$^{-2}$ within a radius of about $1.2\,$pc. Yet closer in, we find 8 X-ray sources within the outermost contour of the $850\,\mu$m map of Cesaroni et al. (1999), i.e. the X-ray source surface density within a distance of $0.75\,$pc, continues to rise to a value of $10.7\,$pc$^{-2}$. Finally, within the typical radius of clusters around Be type stars of about $0.21\,$pc (Testi et al. 1999), we find 3 X-ray sources, corresponding to an X-ray source surface density of $21.4\,$pc$^{-2}$. Hence there is a clear trend of increasing stellar surface density toward the massive protostar.

In conclusion, the massive protostar IRAS$\,20126+4104$ is associated with about 90 X-ray sources, which are likely lower mass YSOs. Most of the X-ray sources are distributed in an extended pc-size halo, and 26 X-rays sources are located within a projected radius of $1.2\,$pc from the massive protostar. These results indicate that an early B-type star is forming in a small cluster. However, mostly due to the limiting sensitivity of our X-ray data these numbers are lower limits, and to determine the richness of the young cluster around the massive protostar IRAS$\,20126+4104$ deeper X-ray observations would be highly desirable.

\section{Summary}

We have observed the IRAS$\,20126+4104$ region with {\it Chandra} and the VLA $6\,$cm continuum band. These data are augmented by NIR and optical archival data. We identify $150$ X-ray sources of which $88$ have NIR counterparts. Based on an NIR color-color analysis we estimate that about 90 X-ray sources are associated with this massive star forming region. We detect an increasing surface density of X-ray sources toward the massive protostar and including our radio detections and the mid-IR detections of Qiu et al. (2008), we infer the presence of a cluster of at least 46 YSOs within a distance of $1.2\,$pc from the massive protostar.

\acknowledgements

P.H. acknowledges partial support from NSF grant AST-0908901. This project was also supported by CXC award G03-4018A. C.A. and V.R. are supported by the NRAO Grote Reber Doctoral Fellowship. We would like to thank D. Shepherd for helpful discussion, and Roc Cutri for help with the preparation of Figure~6. This research has made use of the NASA/ IPAC Infrared Science Archive, which is operated by the Jet Propulsion Laboratory, California Institute of Technology, under contract with the National Aeronautics and Space Administration, as well as of the VizieR catalogue access tool, CDS, Strasbourg, France.

\clearpage

\includegraphics{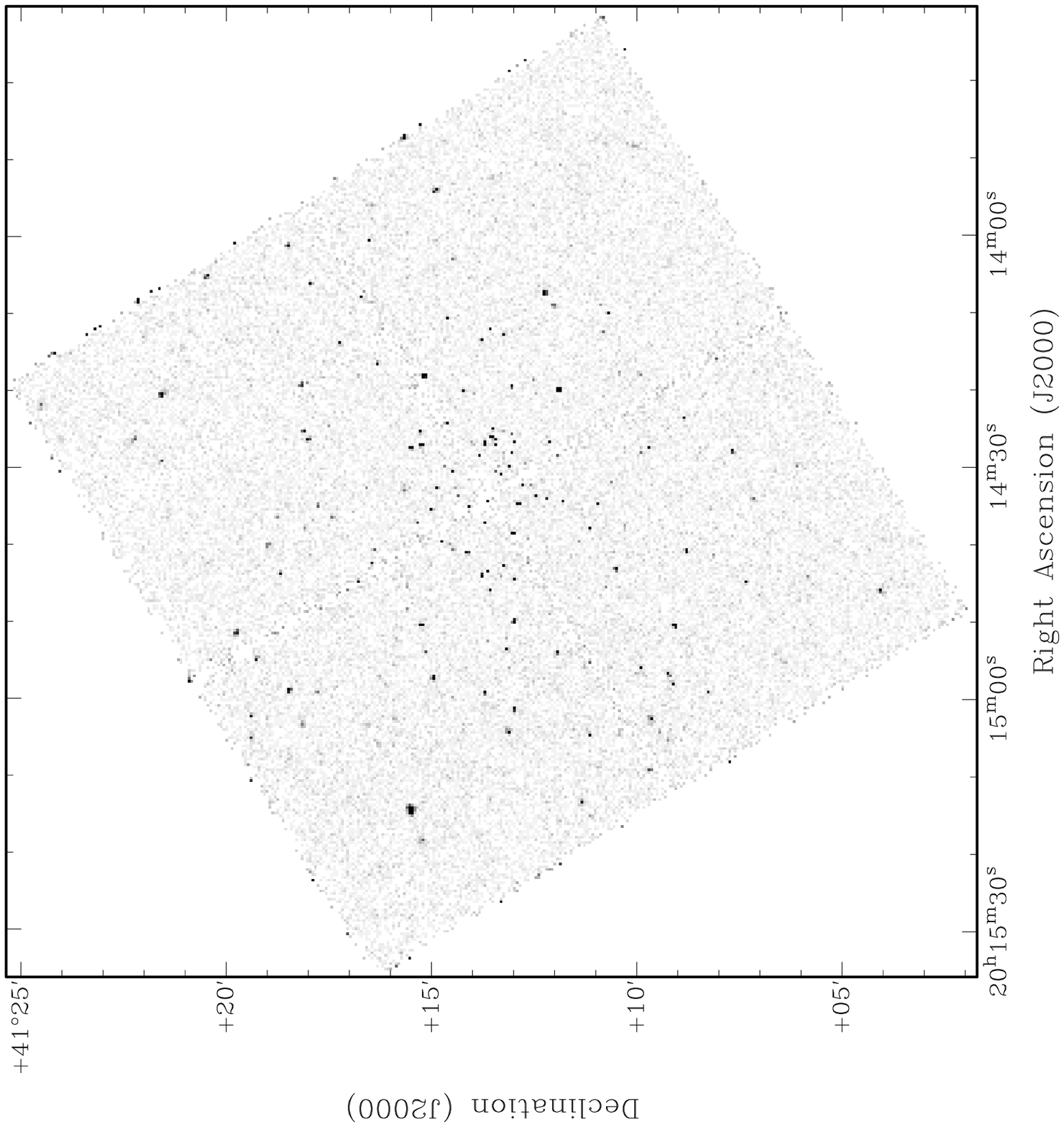}

\hspace*{0.0cm}
\begin{minipage}{15cm}
\vspace{18cm}
{\small Figure~1. Grey scale image of the full $17^\prime \times 17^\prime$ ACIS-I field in the $0.5 - 8\,$keV band toward IRAS$\,20126+4104$.
We detected a total of 150 X-ray point sources.}
\end{minipage}

\clearpage

\includegraphics{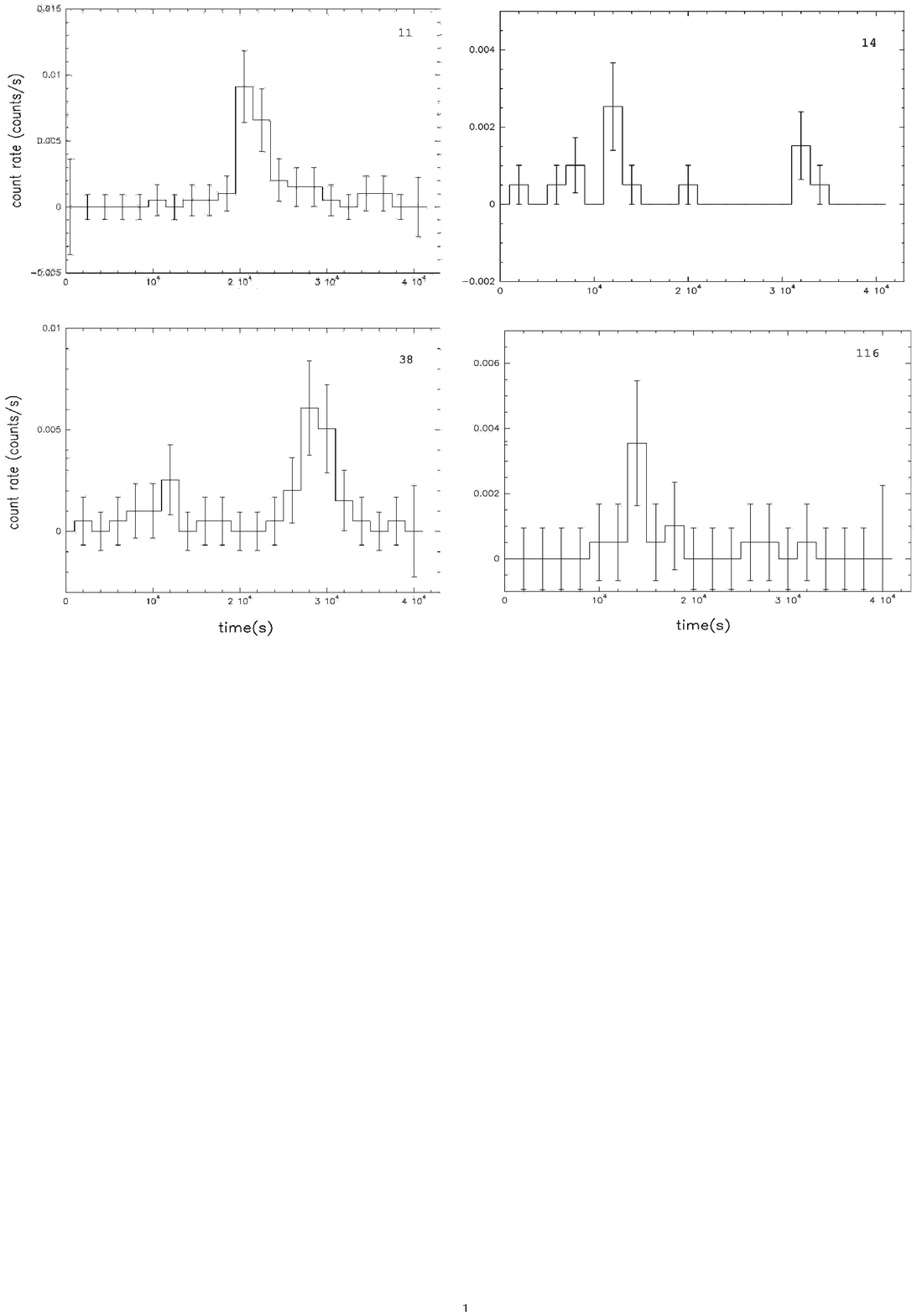}

\hspace*{0cm}
\begin{minipage}{15cm}
\vspace{14cm}
{\small Figure~2. Light curves for sources classified as flare types. Source 11 shows the typical flare behavior with a fast rise and a slow decay. In contrast, source 38 shows a more
symmetric light curve, and  in sources 14 and 116 the flares are unresolved. The multiple peaks in source 14 can possibly be explained with multiple flares during our observations.}
\end{minipage}

\clearpage

\includegraphics{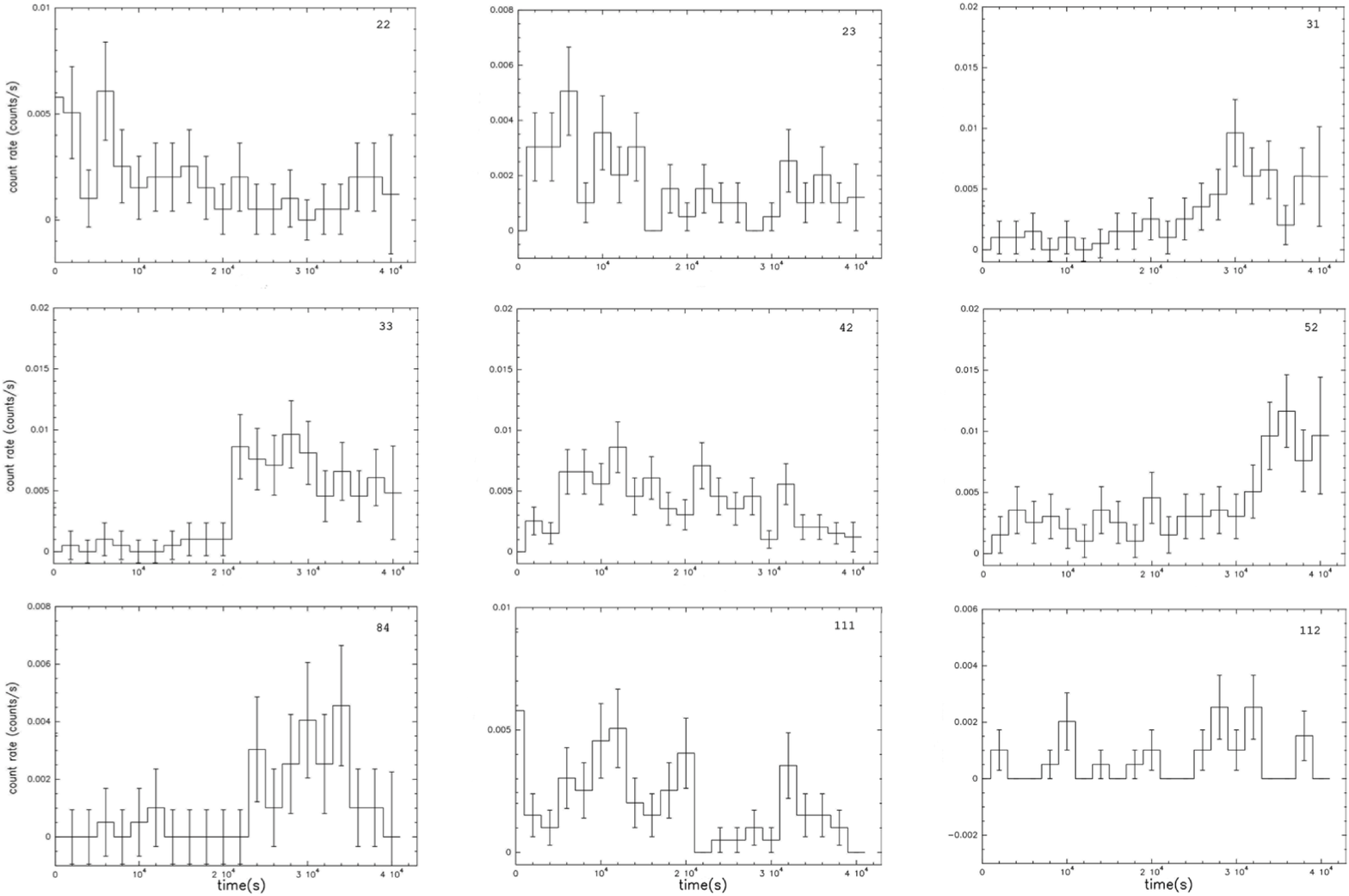}

\hspace*{0.0cm}
\begin{minipage}{15cm}
\vspace{14.5cm}
{\small Figure~3. Light curves for non-flare sources. We can distinguish two types of variability in theses sources. The first type is a slow rise and slow decay (sources 22, 23, 31, 42, 111, 112), the second type is a single step rise or decay (sources 33, 52, 84).}
\end{minipage}

\clearpage

\includegraphics{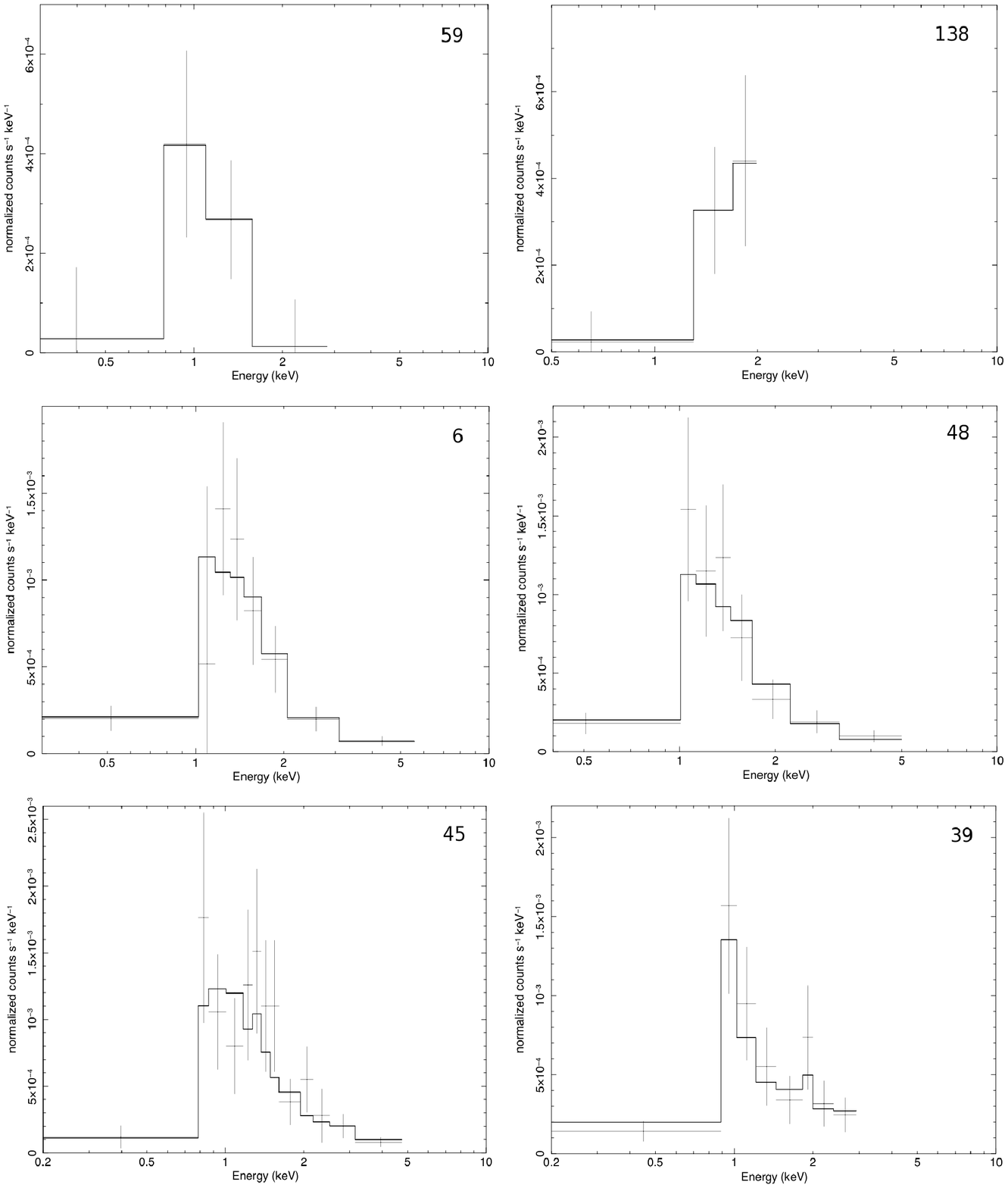}

\hspace*{0.0cm}
\begin{minipage}{15cm}
\vspace{17cm}
{\small Figure~4. Sample spectra for two sources from each photon count group as described in the text. Sources 59 and 138 belong to the first group: low counts, fitted with a simple thermal model and a single absorption component, resulting in good spectral fits but non-unique physical parameters. Sources 6 and 48 belong to the second group:
sufficient photon counts to fit both T and N$_H$ with a simple thermal model and a single absorption component. Sources 45 and 39 belong to the third group: two thermal components with two separate absorption columns.}
\end{minipage}

\clearpage

\includegraphics{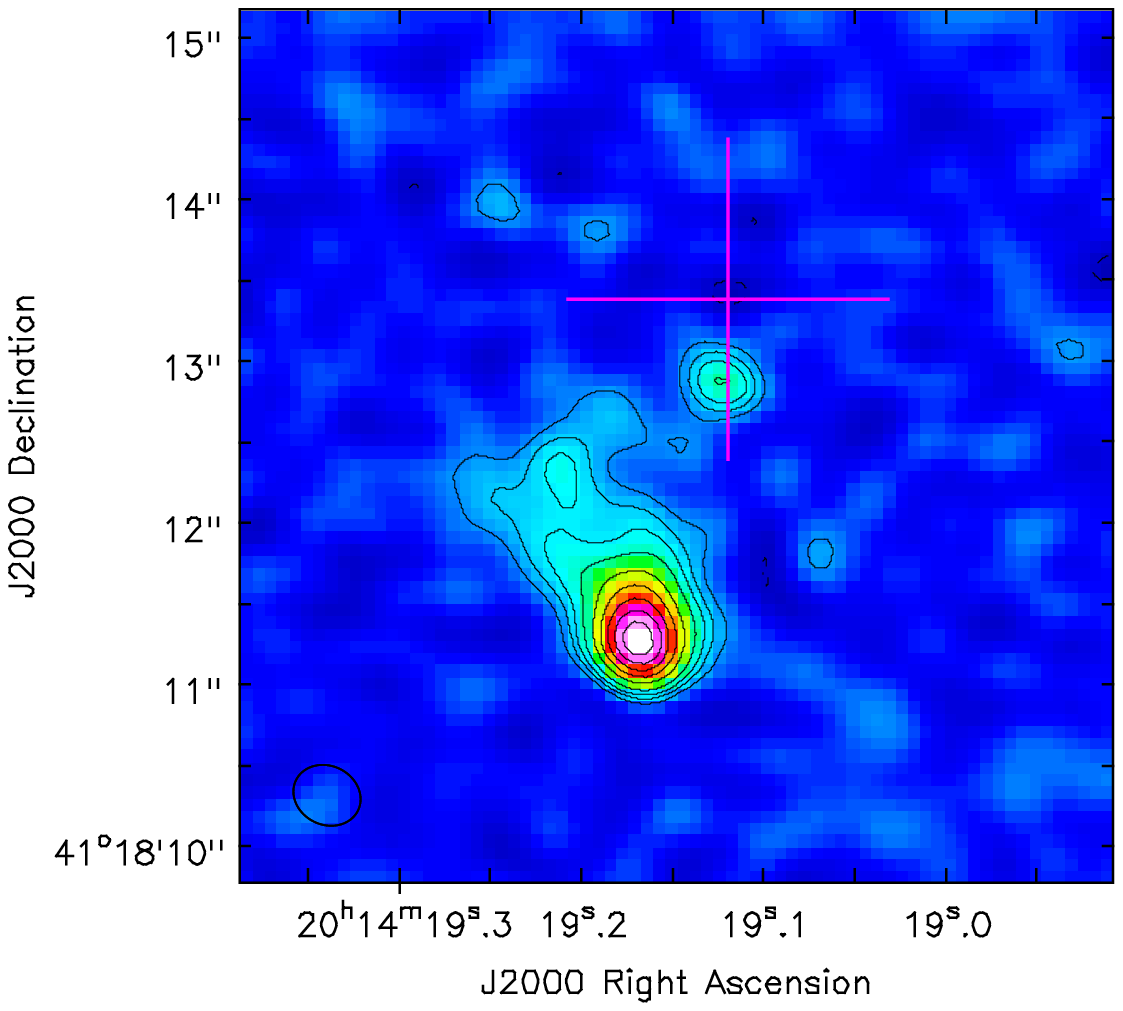}
\includegraphics{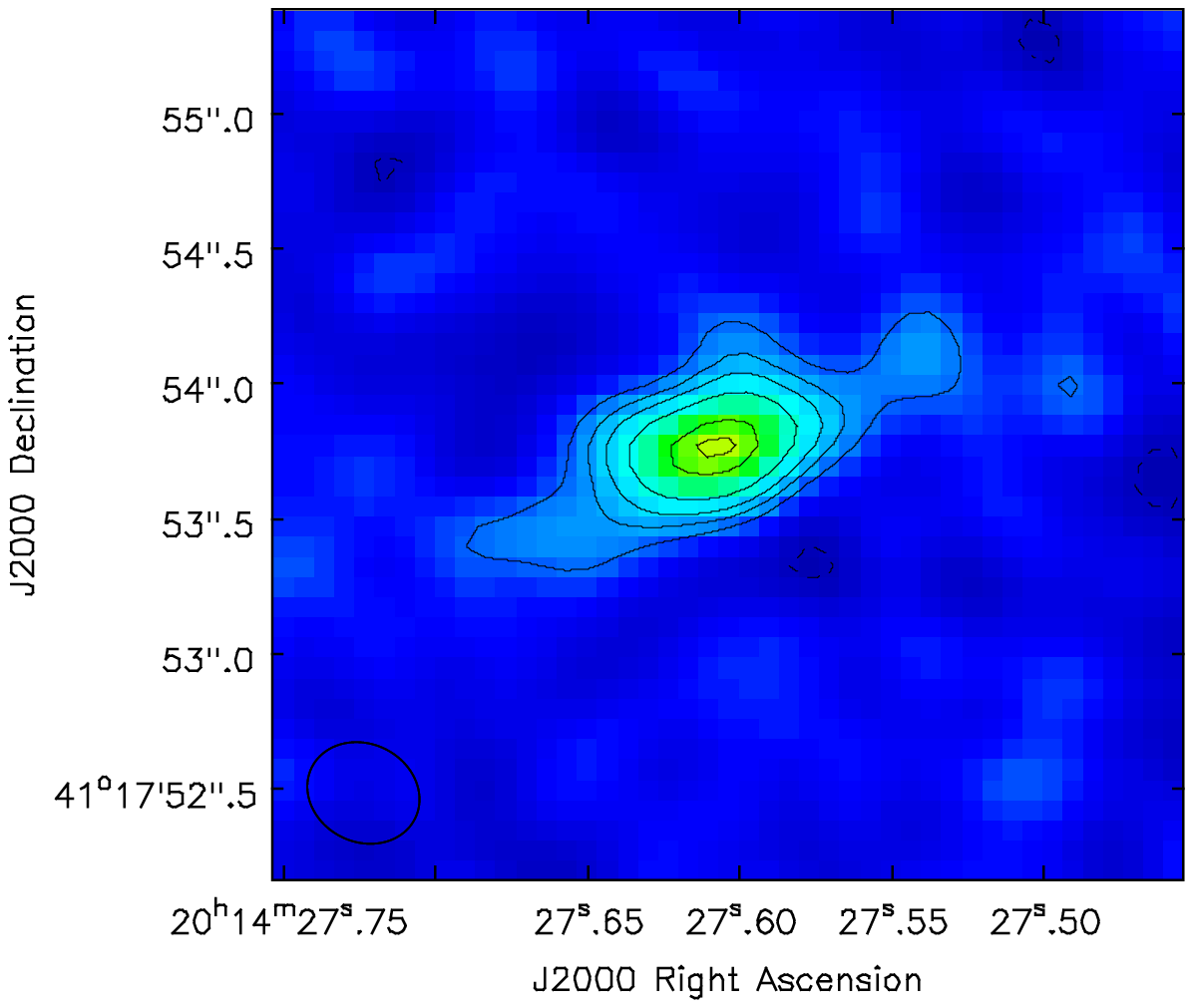}

\hspace*{0.0cm}
\begin{minipage}{15cm}
\vspace{12cm}
{\small Figure~5. $6\,$cm VLA images of extended sources in the IRAS$\,20126+4104$ region. Left panel: We detect one extended radio source (G78.1750+3.6936) and one point source (G78.1790+3.5640). They are both located outside the 4.9 GHz primary beam. Contour levels are -3, 3, 5, 7, 10, 15, 20, 25, 30, 35, 37 $\times$ 19$\,\mu$Jy/b. The point source has an X-ray counterpart (source \#30); its position is marked with a magenta cross. The size of the cross corresponds to the approximate position error of the X-ray counterparts with off-axis position $\theta > 3 \arcmin$. Right panel: This extended source, G78.2537+3.6689,
is located outside the 7.4 primary beam and shows an elongated shape, reminiscent of a spiral galaxy seen edge on.
Contour levels are -3, 3, 5, 7, 10, 15, 17 $\times$ 16$\,\mu$Jy/b}.
\end{minipage}

\clearpage

\includegraphics{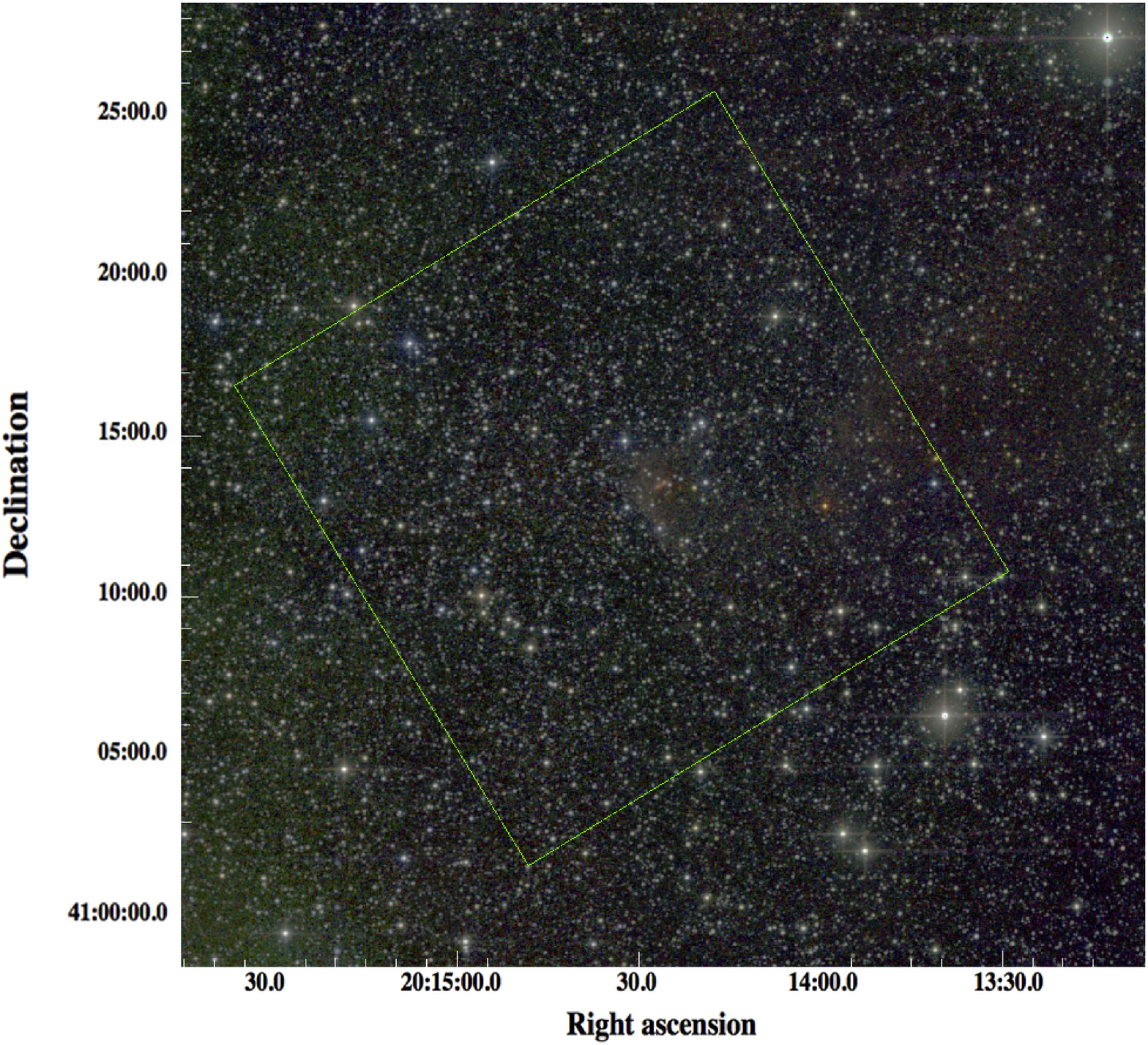}

\hspace*{0cm}
\begin{minipage}{15cm}
\vspace{17.5cm}
{\small Figure~6. $30^\prime \times 30^\prime$ 2MASS three color image with J, H and K$_{s}$ band composite of the IRAS 20126+4104 region, the green box indicates the position of the ACIS 
detector. The image illustrates the difficulties of identifying a cluster associated with the central proto-star in the infrared wavelength region, which are due to foreground/background contamination,
as well as the presence of extended nebulosity. }
\end{minipage}

\clearpage

\includegraphics{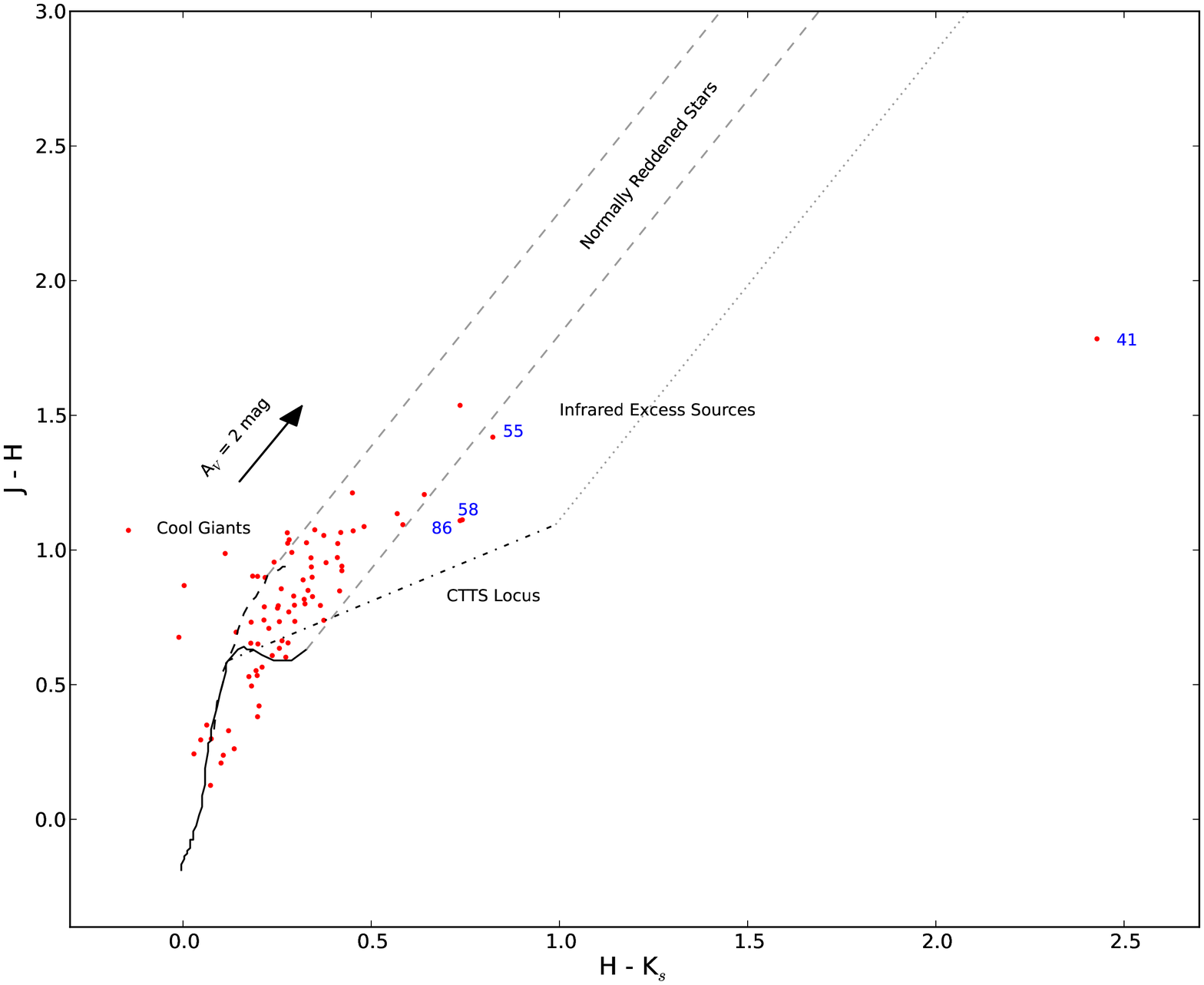}

\hspace*{0.0cm}
\begin{minipage}{15cm}
\vspace{14cm}
{\small Figure~7. Color-color diagram of IRAS$\,$20126+4104 2MASS counterparts. The solid and dashed black lines are the locus for main-sequence and giant branch from Cox (2000). The dashed gray lines denote the reddening band for main sequence and giant colors from Rieke and Lebofsky (1985) starting at M4 III for giants (left line) and M6 V for main-sequence (right line). The dashed-dotted black line denote the locus for CTTS from Meyer et al. (1997) starting at K2 V. The gray dotted line marks the reddening band corresponding to CTTS colors. The sources presenting an infrared excess are numbered in blue. Note that the symbols for sources 58 and 86 overlap in the diagram.}
\end{minipage}

\clearpage

\includegraphics{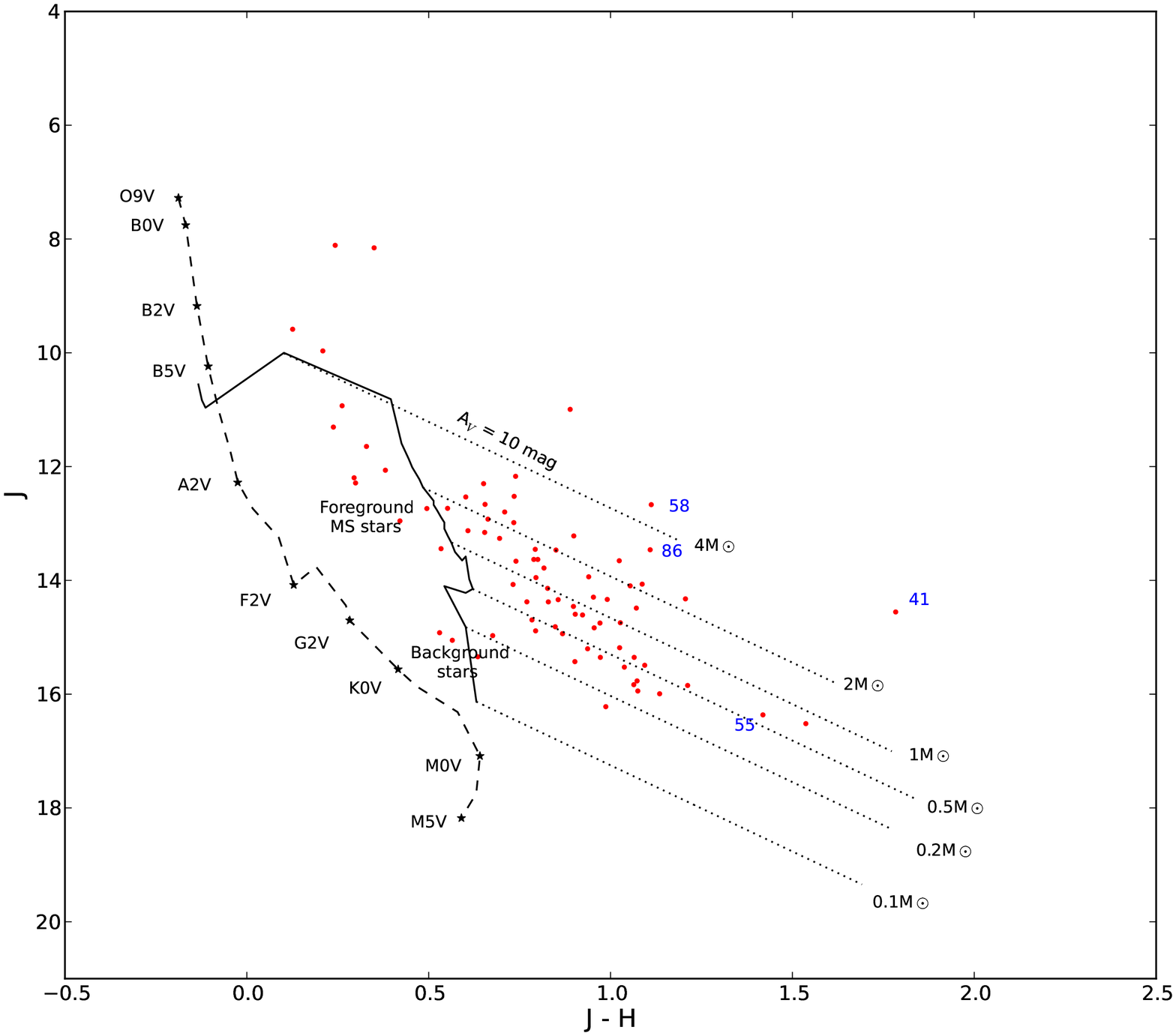}

\hspace*{0.0cm}
\begin{minipage}{15cm}
\vspace{14cm}
{\small Figure~8. J vs. J-H color-magnitude diagram of IRAS$\,$20126+4104 2MASS counterparts. The dashed line marks the location of the unabsorbed main-sequence stars from Cox (2000). Spectral types are marked by black stars. The solid line is the 1 Myr isochrone (Siess et al. 2000), from which A$_V$ $\approx$ 10 mag reddening vectors (dotted lines) are shown for 0.1 M$_{\odot}$, 0.2 M$_{\odot}$, 0.5 M$_{\odot}$, 1 M$_{\odot}$, 2 M$_{\odot}$ and 4 M$_{\odot}$. The unabsorbed main-sequence and the isochrone were corrected for the 2MASS photometry using transformations from Carpenter (2001).}
\end{minipage}

\clearpage

\begin{deluxetable}{ccrrrccc}
	\tabletypesize{\scriptsize}
	\tablecaption{Observed X-Ray Properties}
	\tablenum{1}
	\tablehead{\colhead{}&\colhead{}&\colhead{R.A.}&\colhead{Dec.}&\colhead{CR}&\colhead{F$_{x}$}& \colhead{HR}&\colhead{Var.Cl.}\\ 
\colhead{Source}&\colhead{CXO J}&\colhead{(s)}&\colhead{($^{\prime\prime}$)}&\colhead{(cts ks$^{-1}$)}&\colhead{10$^{-14}$(ergs cm$^{-2}$ s$^{-1}$)}& \colhead{}&\colhead{}\\
\colhead{(1)}&\colhead{(2)}&\colhead{(3)}&\colhead{(4)}&\colhead{(5)}&\colhead{(6)}&\colhead{(7)}&\colhead{(8)}} 

\startdata
1	&	201347.1+411541.3	&	47.079	&	41.290	&	0.88	&	1.43			&	$-$	&				\\
2	&	201347.6+411441.8	&	47.613	&	41.780	&	0.40	&	0.74			&	$-$	&				\\
3	&	201348.1+411046.8	&	48.057	&	46.830	&	0.41	&	1.15			&	$1.00$	&		\\
4	&	201348.3+411007.1	&	48.280	&	 7.070	&	0.33	&	0.77			&	$1.00$	&		\\
5	&	201352.3+411333.6	&	52.273	&	33.640	&	0.32	&	0.32			&	$-1.00$	&			\\
6	&	201353.9+411455.3	&	53.920	&	55.330	&	1.78	&	1.43			&	$-0.58$	&				\\
7	&	201355.8+411629.0	&	55.786	&	29.050	&	0.47	&	1.28			&	$-$	&				\\
8	&	201400.4+411632.8	&	 0.421	&	32.810	&	0.46	&	0.41			&	$-1.00$	&			\\
9	&	201401.0+411830.7	&	 1.049	&	30.680	&	0.71	&	0.37			&	$-1.00$	&			\\
10	&	201402.8+411430.0	&	 2.838	&	30.040	&	0.24	&	0.24			&	$-$	&			\\
11	&	201405.1+412031.2	&	 5.086	&	31.190	&	1.34	&	$-$			&	$-0.86$	&	FL			\\
12	&	201405.9+411758.9	&	 5.879	&	58.860	&	0.33	&	0.23			&	$-1.00$	&			\\
13	&	201407.3+411215.8	&	 7.258	&	15.780	&	1.72	&	2.12			&	$-0.79$	&				\\
14	&	201408.9+411055.9	&	 8.911	&	55.860	&	0.21	&	0.40			&	$-$	&	FL			\\
15	&	201408.9+411203.7	&	 8.913	&	 3.670	&	0.42	&	0.52			&	$-1.00$	&			\\
16	&	201409.9+411043.7	&	 9.924	&	43.740	&	0.47	&	0.58			&	$1.00$	&			\\
17	&	201410.5+411439.1	&	10.464	&	39.080	&	0.52	&	0.32			&	$-1.00$	&			\\
18	&	201411.3+410734.9	&	11.334	&	34.860	&	0.25	&	0.94			&	$-$	&				\\
19	&	201412.0+411336.9	&	11.951	&	36.940	&	0.31	&	0.21			&	$-$	&				\\
20	&	201412.2+412056.2	&	12.161	&	56.180	&	0.28	&	5.05			&	$1.00$	&			\\
21	&	201412.5+411052.0	&	12.451	&	51.990	&	0.62	&	$-$			&	$-$	&				\\
22	&	201412.6+411316.0	&	12.601	&	15.980	&	1.99	&	$-$			&	$-0.41$	&	V			\\
23	&	201413.5+411348.9	&	13.466	&	48.850	&	2.07	&	1.87			&	$-0.42$	&	V			\\
24	&	201413.7+411715.5	&	13.737	&	15.500	&	0.23	&	0.29			&	$-1.00$	&			\\
25	&	201414.1+412106.9	&	14.051	&	 6.930	&	0.24	&	0.30			&	$-$	&				\\
26	&	201416.3+411620.5	&	16.332	&	20.520	&	0.73	&	0.38			&	$-1.00$	&			\\
27	&	201417.8+410957.6	&	17.770	&	57.640	&	0.46	&	0.52			&	$-$	&				\\
28	&	201418.1+411512.7	&	18.059	&	12.690	&	1.66	&	1.52			&	$-0.15$	&				\\
29	&	201418.4+411231.8	&	18.446	&	31.790	&	0.36	&	0.53			&	$1.00$	&			\\
30	&	201419.1+411813.4	&	19.116	&	13.390	&	0.45	&	0.70			&	$-$	&				\\
31	&	201419.4+411304.9	&	19.412	&	 4.880	&	2.88	&	$-$			&	$-0.41$	&	V			\\
32	&	201419.8+411155.9	&	19.829	&	55.940	&	2.18	&	1.33			&	$-0.51$	&				\\
33	&	201420.0+411415.3	&	19.986	&	15.300	&	3.63	&	12.90			&	$-0.35$	&	V		\\
34	&	201420.4+412136.9	&	20.392	&	36.920	&	2.06	&	$-$		&		$0.38$	&				\\
35	&	201421.8+412432.6	&	21.844	&	32.570	&	1.13	&	1.61			&	$0.15$	&			\\
36	&	201423.5+410853.3	&	23.470	&	53.310	&	0.31	&	0.21			&	$-1.00$	&			\\
37	&	201425.0+411332.0	&	24.993	&	31.960	&	1.78	&	4.04			&	$-0.38$	&				\\
38	&	201425.2+411809.1	&	25.233	&	 9.140	&	1.22	&	0.67			&	$-0.03$	&	FL			\\
39	&	201425.3+411517.9	&	25.342	&	17.930	&	1.42	&	2.49	&	$-0.40$	&				\\		
40	&	201425.6+411305.7	&	25.601	&	 5.720	&	0.27	&	0.81			&	$-$	&				\\
41	&	201425.9+411336.5	&	25.861	&	36.510	&	0.49	&	0.66			&	$-1.00$	&			\\
42	&	201426.0+411331.7	&	26.031	&	31.740	&	4.29	&	13.0			&	$1.00$	&	V		\\
43	&	201426.2+411327.9	&	26.245	&	27.900	&	0.85	&	2.70			&	$1.00$	&			\\
44	&	201426.3+412218.2	&	26.315	&	18.200	&	0.87	&	1.17			&	$1.00$	&			\\
45	&	201426.3+411802.7	&	26.334	&	 2.720	&	1.86	&	3.38  	&	$-0.41$	&				\\		
46	&	201426.4+412403.1	&	26.423	&	 3.120	&	0.44	&	$-$			&	$-1.00$	&			\\
47	&	201426.6+411210.2	&	26.583	&	10.160	&	0.38	&	0.49			&	$-1.00$	&			\\
48	&	201426.7+411300.9	&	26.704	&	 0.880	&	1.57	&	1.38			&	$-0.38$	&				\\
49	&	201426.8+411343.0	&	26.808	&	42.970	&	2.00	&	 2.80 	&	$-0.57$	&				\\		
50	&	201427.0+411328.7	&	26.996	&	28.710	&	2.07	&	 4.46 	&	$-0.24$	&				\\		
51	&	201427.1+411516.8	&	27.114	&	16.780	&	0.81	&	0.67			&	$-1.00$	&			\\
52	&	201427.4+411533.1	&	27.374	&	33.120	&	4.29	&	9.73  	&	$-0.90$	&	V			\\		
53	&	201427.4+410944.4	&	27.393	&	44.390	&	0.48	&      2.04			&	$-1.00$	&			\\
54	&	201427.8+410741.4	&	27.839	&	41.380	&	0.84	&	1.00			&	$0.31$	&				\\
55	&	201428.0+411303.4	&	27.953	&	 3.410	&	0.32	&	0.39			&	$1.00$	&			\\
56	&	201428.1+410954.2	&	28.090	&	54.240	&	0.33	&	0.33			&	$-$	&				\\
57	&	201428.2+411317.3	&	28.173	&	17.320	&	0.25	&	0.18			&	$-$	&				\\
58	&	201428.4+411351.8	&	28.387	&	51.820	&	1.96	&	3.21	&	$-0.37$	&				\\		
59	&	201429.0+412137.1	&	29.046	&	37.140	&	0.52	&      0.25			&	$-1.00$	&			\\
60	&	201429.6+410607.9	&	29.600	&	 7.890	&	0.35	&	0.34			&	$-$	&				\\
61	&	201429.7+411444.2	&	29.715	&	44.230	&	0.25	&	0.28			&	$-1.00$	&			\\
62	&	201429.8+411309.2	&	29.773	&	 9.210	&	1.01	&	1.00			&	$-0.58$	&				\\
63	&	201429.8+411322.8	&	29.782	&	22.780	&	0.24	&	0.14			&	$-$	&				\\
64	&	201430.3+411328.7	&	30.342	&	28.670	&	0.27	&	0.21			&	$-1.00$	&			\\
65	&	201430.4+411430.7	&	30.392	&	30.730	&	1.42	&	0.84			&	$-1.00$	&			\\
66	&	201430.9+411321.9	&	30.856	&	21.900	&	0.38	&	0.38			&	$-1.00$	&			\\
67	&	201431.6+411219.5	&	31.647	&	19.510	&	0.25	&	0.22			&	$-1.00$	&			\\
68	&	201431.8+411038.3	&	31.782	&	38.260	&	0.20	&	0.56			&	$-$	&				\\
69	&	201432.1+411249.6	&	32.073	&	49.560	&	0.75	&	0.81			&	$-1.00$	&			\\
70	&	201432.5+411453.8	&	32.453	&	53.810	&	3.60	&	$-$			&	$-1.00$	&			\\
71	&	201432.7+411544.0	&	32.750	&	44.000	&	0.36	&	0.63			&	$-1.00$	&			\\
72	&	201433.6+411230.5	&	33.583	&	30.490	&	0.86	&	0.50			&	$-1.00$	&			\\
73	&	201433.7+411424.4	&	33.656	&	24.420	&	0.26	&	0.42			&	$1.00$	&			\\
74	&	201433.9+410712.5	&	33.903	&	12.480	&	0.28	&	0.15			&	$-$	&				\\
75	&	201434.0+411212.9	&	33.968	&	12.900	&	0.21	&	0.38			&	$-1.00$	&			\\
76	&	201434.3+411150.3	&	34.335	&	50.270	&	1.57	&	0.95			&	$-0.57$	&				\\
77	&	201434.5+411255.0	&	34.529	&	55.020	&	1.05	&	0.59			&	$-0.60$	&				\\
78	&	201434.6+411059.8	&	34.650	&	59.820	&	0.30	&	0.26			&	$-$	&				\\
79	&	201434.8+411748.6	&	34.846	&	48.630	&	0.42	&	0.56			&	$-1.00$	&			\\
80	&	201435.3+411502.5	&	35.341	&	 2.510	&	0.30	&	0.43			&	$1.00$	&			\\
81	&	201436.2+411726.5	&	36.249	&	26.530	&	0.43	&	0.25			&	$-1.00$	&			\\
82	&	201436.3+411848.5	&	36.269	&	48.500	&	0.33	&	0.40			&	$-$	&				\\
83	&	201437.1+411521.6	&	37.128	&	21.640	&	0.29	&	0.23			&	$-$	&				\\
84	&	201437.1+411343.3	&	37.142	&	43.340	&	1.23	&	1.28			&	$-0.14$	&	V			\\
85	&	201437.5+411020.4	&	37.537	&	20.430	&	0.27	&	0.31			&	$-$	&				\\
86	&	201437.9+411111.8	&	37.871	&	11.780	&	0.52	&	0.25			&	$-1.00$	&			\\
87	&	201438.4+411302.6	&	38.402	&	 2.620	&	1.38	&	0.73			&	$-1.00$	&			\\
88	&	201439.4+411125.5	&	39.394	&	25.470	&	0.21	&	0.37			&	$-$	&				\\
89	&	201439.6+411446.3	&	39.562	&	46.350	&	0.68	&	0.36			&	$-1.00$	&			\\
90	&	201440.0+411901.5	&	40.025	&	 1.490	&	0.62	&	0.41			&	$-1.00$	&			\\
91	&	201440.4+411429.7	&	40.411	&	29.660	&	0.27	&	0.27			&	$-$	&				\\
92	&	201440.6+411627.2	&	40.554	&	27.240	&	0.29	&	0.17			&	$-1.00$	&			\\
93	&	201440.8+410849.6	&	40.758	&	49.590	&	1.34	&	1.78			&	$0.24$	&				\\
94	&	201440.9+411037.0	&	40.915	&	37.000	&	0.26	&      0.12			&	$-1.00$	&			\\
95	&	201441.0+411409.8	&	41.028	&	 9.800	&	0.68	&	0.36			&	$-1.00$	&			\\
96	&	201441.2+410443.4	&	41.240	&	43.390	&	0.38	&	0.31			&	$-1.00$	&			\\
97	&	201442.5+412031.0	&	42.474	&	31.040	&	0.27	&	0.32			&	$-1.00$	&			\\
98	&	201442.7+411315.5	&	42.711	&	15.470	&	1.34	&	$-$			&	$-0.07$	&				\\
99	&	201443.1+411032.3	&	43.080	&	32.350	&	2.14	&	$-$			&	$0.23$	&				\\
100	&	201443.4+411340.5	&	43.395	&	40.510	&	0.33	&	0.25			&	$-1.00$	&			\\
101	&	201443.7+411843.4	&	43.671	&	43.380	&	0.91	&	1.26			&	$-0.15$	&				\\
102	&	201443.9+411348.6	&	43.910	&	48.640	&	0.73	&	0.80			&	$1.00$	&			\\
103	&	201444.5+411259.6	&	44.460	&	59.620	&	0.38	&	0.42			&	$-1.00$	&			\\
104	&	201444.8+410721.8	&	44.772	&	21.820	&	0.93	&	0.52			&	$-1.00$	&			\\
105	&	201445.3+411539.0	&	45.278	&	39.010	&	0.22	&	0.17			&	$-$	&				\\
106	&	201445.9+411337.5	&	45.879	&	37.490	&	1.10	&	0.60			&	$-1.00$	&			\\
107	&	201445.9+410404.9	&	45.898	&	 4.930	&	1.60	&	1.19			&	$-0.79$	&				\\
108	&	201449.9+411302.0	&	49.897	&	 2.000	&	2.12	&	1.57			&	$-0.67$	&				\\
109	&	201450.3+410547.2	&	50.282	&	47.190	&	0.25	&	0.72			&	$-1.00$	&			\\
110	&	201450.3+411517.1	&	50.316	&	17.070	&	2.17	&	2.08			&	$-0.61$	&				\\
111	&	201450.4+410906.9	&	50.416	&	 6.920	&	2.03	&	1.31			&	$-1.00$	&	V		\\
112	&	201451.4+411948.0	&	51.393	&	48.030	&	0.85	&	1.03			&	$-1.00$	&	V		\\
113	&	201452.4+411340.8	&	52.439	&	40.800	&	0.21	&	0.23			&	$-$	&				\\
114	&	201453.5+411312.8	&	53.475	&	12.780	&	0.79	&	0.79			&	$-0.41$	&				\\
115	&	201453.8+411158.6	&	53.822	&	58.640	&	0.99	&	0.70			&	$-0.27$	&				\\
116	&	201453.9+411512.1	&	53.906	&	12.070	&	0.27	&	0.27			&	$-1.00$	&	FL		\\
117	&	201454.7+411917.5	&	54.700	&	17.530	&	0.80	&	0.74		        &	$1.00$	&			\\
118	&	201455.9+410955.6	&	55.862	&	55.560	&	0.58	&	0.53			&	$0.21$	&				\\
119	&	201456.7+410915.2	&	56.734	&	15.150	&	0.92	&	1.24			&	$-0.25$	&				\\
120	&	201457.0+411618.8	&	57.000	&	18.800	&	0.31	&	 0.55			&	$-1.00$	&			\\
121	&	201457.3+411459.3	&	57.286	&	59.290	&	2.43	&	 1.01			&	$0.19$	&				\\
122	&	201457.9+410909.5	&	57.874	&	 9.510	&	0.62	&	$-$			&	$-1.00$	&			\\
123	&	201458.9+411829.5	&	58.850	&	29.490	&	1.22	&	 1.58			&	$-0.02$	&				\\
124	&	201459.1+410815.6	&	59.066	&	15.610	&	0.49	&	0.33			&	$0.02$	&				\\
125	&	201459.1+411749.7	&	59.126	&	49.700	&	0.54	&	0.70			&	$-1.00$	&			\\
126	&	201459.2+411343.8	&	59.224	&	43.770	&	0.67	&	0.64			&	$-$	&				\\
127	&	201501.3+411301.2	&	 1.320	&	 1.170	&	0.77	&	0.69			&	$-0.12$	&				\\
128	&	201501.6+410732.7	&	 1.579	&	32.670	&	0.25	&      0.42			&	$1.00$	&			\\
129	&	201502.2+411925.5	&	 2.155	&	25.470	&	0.60	&	0.42			&	$-1.00$	&			\\
130	&	201502.4+410941.1	&	 2.439	&	41.090	&	1.33	&	$-$			&	$-0.48$	&				\\
131	&	201502.7+411556.2	&	 2.681	&	56.170	&	0.28	&	0.29			&	$-1.00$	&			\\
132	&	201503.3+411809.7	&	 3.260	&	 9.720	&	1.02	&	1.44			&	$-0.05$	&				\\
133	&	201503.4+411635.6	&	 3.392	&	35.590	&	0.34	&	0.49			&	$-$	&				\\
134	&	201503.6+410851.9	&	 3.603	&	51.900	&	0.37	&	0.19			&	$-1.00$	&			\\
135	&	201504.1+411309.7	&	 4.133	&	9.700	&	1.02	&	1.45			&	$-0.23$	&				\\
136	&	201504.2+410928.7	&	 4.236	&	28.660	&	0.25	&	0.25			&	$-$	&				\\
137	&	201504.5+411110.1	&	 4.477	&	10.140	&	0.33	&	0.37			&	$-1.00$	&			\\
138	&	201504.5+411026.4	&	 4.519	&	26.400	&	0.37	&	0.36			&	$-1.00$	&			\\
139	&	201504.9+411925.5	&	 4.925	&	25.500	&	1.02	&	1.61			&	$1.00$	&			\\
140	&	201505.8+411619.3	&	 5.819	&	19.320	&	0.39	&	0.40			&	$-1.00$	&			\\
141	&	201505.2+410914.6	&	 5.246	&	14.580	&	0.52	&	0.40			&	$-$	&				\\
142	&	201507.2+410834.7	&	 7.200	&	34.660	&	0.28	&	0.75			&	$-$	&				\\
143	&	201508.8+411337.8	&	 8.780	&	37.760	&	0.24	&	0.21			&	$-$	&				\\
144	&	201509.1+410945.0	&	 9.057	&	45.000	&	0.65	&	0.45			&	$-1.00$	&			\\
145	&	201510.1+411439.7	&	10.088	&	39.740	& 	0.21	&	0.26			&	$-1.00$	&			\\
146	&	201513.0+411336.7	&	13.044	&	36.710	&	0.37	&	0.25			&	$-1.00$	&			\\
147	&	201513.2+411121.9	&	13.226	&	21.930	&	0.66	&	0.47			&	$-1.00$	&			\\
148	&	201514.4+411531.9	&	14.390	&	31.890	&	9.90	&	$-$			&	$-1.00$	&			\\
149	&	201515.3+411107.4	&	15.257	&	 7.410	&	0.41	&	0.33			&	$-$	&				\\
150	&	201518.4+411515.6	&	18.430	&	15.610	&	1.13	&	2.21  		&	$1.00$	&			\\

\hline
\enddata

\tablecomments{
In the variable classification column the source is either 
V Variable, FL Flaring, or constant if left blank.
No entry in the F$_x$ column indicates that no acceptable model fit could be found.
No entry in the hardness ratio column (HR) indicates that the source was only detected
in the full energy range.}

\end{deluxetable}

\clearpage

\begin{deluxetable}{ccccccc}
	\tabletypesize{\scriptsize}
	\tablecolumns{7}
    \tablewidth{0pc}
	\tablecaption{X-ray sources fitted with the 1T models: wabs $\times$ apec}
	\tablenum{2}
%	\rotate

	\tablehead{\colhead{}&\colhead{}&\colhead{CR}&\colhead{kT}&\colhead{N$_H$}&\colhead{F$_{x,c}$}&\colhead{log L$_{x,c}$}\\
\colhead{Source}&\colhead{CXO J}&\colhead{(cts ks$^{-1}$)}&\colhead{(keV)}&\colhead{10$^{22}$ (cm$^{-2}$)}&\colhead{10$^{-13}$ (ergs cm$^{-2}$s$^{-1}$)}&\colhead{(ergs s$^{-1}$)}}  

\startdata
\vspace{0.15cm}
6           & 201353.9+411455.3	& 1.78 & 2.63$^{+1.65}_{-0.79}$ & 0.22$^{+0.20}_{-0.15}$ & 0.1910	&	30.82  \\
\vspace{0.15cm}
13         &	 201407.3+411215.8	& 1.72 & 1.84$^{+1.34}_{-0.50}$ & 2.14$^{+0.55}_{-0.46}$ & 0.6840	&  31.38	  \\
\vspace{0.15cm}
23         &	 201413.5+411348.9	& 2.07 & 3.44$^{+2.68}_{-1.06}$ & 0.19$^{+0.17}_{-0.13}$ & 0.2300   &	 30.90  \\
\vspace{0.15cm}
32         & 201419.8+411155.9	& 2.18 & 0.97$^{+0.29}_{-0.20}$ & 0.76$^{+0.17}_{-0.14}$ & 0.4590	& 31.20	  \\
\vspace{0.15cm}
48         & 201426.7+411300.9	& 1.57 & 2.90$^{+2.94}_{-1.25}$ & 0.22$^{+0.22}_{-0.14}$ & 0.1760	&	30.79  \\
\vspace{0.15cm}
65         &	 201430.4+411430.7	& 1.42 & 0.82$^{+0.24}_{-0.19}$ & 0.82$^{+0.19}_{-0.17}$ & 0.3710	&	31.11   \\
\vspace{0.15cm}
76         &	 201434.3+411150.3	& 1.57 & 0.92$^{+0.28}_{-0.24}$ & 1.04$^{+0.37}_{-0.30}$ & 0.5660	&	31.29   \\
\vspace{0.15cm}
108       &	 201449.9+411302.0	& 2.12 & 1.18$^{+0.28}_{-0.30}$ & 1.16$^{+0.44}_{-0.36}$ & 0.5460	&  31.28	  \\
\vspace{0.15cm}
110       &	 201450.3+411517.1	& 2.17 & 2.98$^{+3.47}_{-1.71}$ & 0.35$^{+0.35}_{-0.22}$ & 0.3140	&	31.04   \\
\vspace{0.15cm}
111\tablenotemark{a}        &	 201450.4+410906.9	& 2.03 & 0.18$^{+0.10}_{-0.06}$ & 0.70$^{+0.14}_{-0.12}$ & 5.8600	& $-$   \\
\enddata
 \tablenotetext{a}{Source 111 has NIR colors consistent with a lightly reddened main sequence star and is hence likely located in
 the foreground.}
 
\end{deluxetable}

\clearpage

\begin{deluxetable}{cccccccc}
	\tabletypesize{\scriptsize}
	\tablecolumns{7}
    \tablewidth{0pc}
	\tablecaption{X-ray sources fitted with 2T models: wabs $\times$ apec + wabs $\times$ apec}
	\tablenum{3}
%	\rotate

	\tablehead{\colhead{}&\colhead{}&\colhead{CR}&\colhead{kT}&\colhead{N$_H$}&\colhead{F$_{x,c}$}&\colhead{log L$_{x,c}$}\\
\colhead{Source}&\colhead{CXO J}&\colhead{(cts ks$^{-1}$)}&\colhead{(keV)}&\colhead{10$^{22}$ (cm$^{-2}$)}&\colhead{10$^{-13}$ (ergs cm$^{-2}$s$^{-1}$)}&\colhead{(ergs s$^{-1}$)}}  

\startdata
\vspace{0.15cm}
39  \tablenotemark{a}       &	 201425.3+411517.9 & 1.42 & 0.85$^{+0.33}_{-0.31}$ & 0.32$^{+0.16}_{-0.26}$  & 0.1596 &	30.74   \\
\vspace{0.15cm}
             &                                 	&         & 0.82$^{+1.34}_{-0.27}$ & 5.17$^{+3.38}_{-1.78}$  &  5.059  &  32.24 \\
\vspace{0.15cm}
45 \tablenotemark{b}        & 201426.3+411802.7  & 1.86 & 0.21$^{+0.41}_{-0.08}$ & 1.07$^{+0.54}_{-0.80}$ &   7.29	& $-$  \\
\vspace{0.15cm}
             &                                 	&         & 1.40$^{+3.59}_{-0.34}$ & 3.69$^{+4.20}_{-1.14}$   & 1.625 &  31.75 \\
\vspace{0.15cm}
49 \tablenotemark{a}        & 201426.8+411343.0  & 2.00 & 0.28$^{+0.14}_{-0.12}$ & 2.34$^{+1.29}_{-0.72}$  & 30.31	& 33.02	  \\
\vspace{0.15cm}
             &                                 	&         & 2.21$^{+2.85}_{-0.81}$ & 0.26$^{+0.19}_{-0.14}$  & 0.2357  & 30.91  \\
\vspace{0.15cm}
58 \tablenotemark{c}        &	 201428.4+411351.8 & 1.96 & 1.28$^{+0.54}_{-0.58}$ & 0.32$^{+0.18}_{-0.11}$  & 0.213 & 30.87	   \\
\vspace{0.15cm}
             &                                 	&         & 1.25$^{+0.72}_{-0.57}$ & 2.43$^{+2.81}_{-1.07}$  & 1.171	&  31.61  \\ \enddata
\tablenotetext{a}{Sources 39 and 49 have NIR colors consistent with the WTTS locus.}
\tablenotetext{b}{Source 45 has NIR colors consistent with a lightly reddened main sequence star and is hence likely located in the foreground.}
\tablenotetext{c}{Source 58 has NIR colors consistent with the CTTS locus.}

\end{deluxetable}

\clearpage

\begin{deluxetable}{ccccccc}
	\tabletypesize{\scriptsize}
	\tablecolumns{7}
    \tablewidth{0pc}
	\tablecaption{Radio Sources in the $6\,$cm Continuum Band}
	\tablenum{4}
	
	\tablehead{\colhead{Source}&\colhead{RA}&\colhead{Dec}&\colhead{rms}&\colhead{S$_{\nu}$}&\colhead{radio source}&\colhead{X-ray counterpart}\\
\colhead{}&\colhead{(J2000.0)}&\colhead{(J2000.0)}&\colhead{($\mu$Jy)}&\colhead{($\mu$Jy)}&\colhead{}&\colhead{}}

\startdata
\vspace{0.15cm}
   G78.1907+3.634     & 20:14:25.83	& +41:13:35.9 & 6.3 & 176.3  &   &   \\
\vspace{0.15cm}
   G78.2030+3.6117     & 20:14:33.39  & +41:13:28.5 & 7.0 &  46.0   &   &    \\
\vspace{0.15cm}
   G78.1911+3.6298   & 20:14:27.00 & +41:13:28.7 & 6.3 & 42.4  &   &  50  \\
\vspace{0.15cm}
   G78.1896+3.6317   & 20:14:26.23  & +41:13:28.1 & 6.0 & 235.6  & Variable source\tablenotemark{c} / I20Var\tablenotemark{d}  &  43  \\
\vspace{0.15cm}
   G78.1903+3.6329   & 20:14:26.03  & +41:13:32.6 & 6.2 & 50.9	&  N1\tablenotemark{c} / I20 N1\tablenotemark{d}  &   \\
\vspace{0.15cm}
   G78.1901+3.6329   & 20:14:26.02 	& +41:13:31.7 & 6.2 & 47.3    & S\tablenotemark{c} / I20 S\tablenotemark{d} & 42  \\
\vspace{0.15cm}
   G78.1000+3.6027   & 20:14:30.18  & +41:11:26.1 & 8.4 & 42.9  &   &   \\
\vspace{0.15cm}
      G78.1767+3.6811\tablenotemark{a}   & 20:14:22.79  & +41:17:51.4 & 17.5 & 238.4  &   &   \\
\vspace{0.15cm}
   G78.2537+3.6689\tablenotemark{a}   &	20:14:27.61  & +41:17:53.8 & 17.9 & 657.0	&   &   \\
\vspace{0.15cm}
   G78.1790+3.5640\tablenotemark{a}      &  20:14:42.38 & +41:10:41.7 & 17.2 & 897.6  &   &   \\

\vspace{0.15cm}
   G78.2007+3.6830\tablenotemark{b}       &  20:14:26.50 & +41:19:07.0 & 27.1 &   661.2 &   &  \\
\vspace{0.15cm}
   G78.1750+3.6936\tablenotemark{b}       &  20:14:19.17 & +41:18:11.3 & 19.6 &  1682.0  &   &  \\
\vspace{0.15cm}
   G78.1790+3.5640\tablenotemark{b}       &  20:14:19.12 & +41:18:12.8 & 19.6 & 115.6  &   & 30  \\
   
   \enddata
   \tablenotetext{a}{Sources located outside the $7.4\,$GHz primary beam. Fluxes were measured in the 4.9 GHz map.}
   \tablenotetext{b}{Sources located outside the $4.9\,$GHz primary beam. Fluxes were measured in the 4.9 GHz map.}
    \tablenotetext{c}{See Hofner et al. 2007}
    \tablenotetext{d}{See Anderson et al. 2011}
\end{deluxetable}

\clearpage

\begin{deluxetable}{lcccccccccc}
	\tabletypesize{\scriptsize}
	\tablecolumns{11}
    \tablewidth{0pc}
	\tablecaption{Multiwavelength Counterparts}
	\tablenum{5}
	\rotate

	\tablehead{
	\multicolumn{2}{c}{X-Ray}    & \colhead{}    &
\multicolumn{4}{c}{Infrared/Radio/Optical Stars}  & \colhead{}    &
\multicolumn{3}{c}{2MASS Photometry (mag)} \\
\cline{1-2} \cline{4-7} \cline{9-11}\\ 
\colhead{Source}&\colhead{CXO J}&\colhead{}&\colhead{2MASS J}&\colhead{SSTSLP J}&\colhead{VLA}&\colhead{USNO-B1.0}&\colhead{}&\colhead{J}&\colhead{H}&\colhead{K$_{s}$}}  

\startdata
1      &	201347.1+411541.3  &  	&  201346.92+411541.7   & \nodata  & \nodata  & \nodata &  & 14.33   &	13.12  &	12.48 	 \\
4      &	201348.3+411007.1 	&  	&   \nodata   & \nodata  & \nodata  & 1311-0384325  &  &  \nodata  &	 \nodata  &	 \\
5      &	201352.3+411333.6	&  	&  201352.25+411333.2  &  \nodata  &  \nodata &  1312-0382643 &  & 15.05   & 14.49  & 14.28    \\
6      &	201353.9+411455.3	&  	&  201353.92+411455.5  &  \nodata  & \nodata  & \nodata  &  & 13.66   & 12.63  & 12.22  	  \\
8      &	201400.4+411632.8	&	&   201400.39+411632.3  &  201400.40+411632.3  & \nodata  & 1312-0382708  &  & 14.97  & 14.30  & 14.31  	   \\
9      & 	201401.0+411830.7	&	&   201401.03+411830.9  &  201401.04+411830.8  & \nodata  & 1313-0383405  &  & 11.00  & 10.11  & 9.79     \\
10    &	201402.8+411430.0	& 	&   201402.75+411428.8  & \nodata  & \nodata &  1312-0382742 &  & 14.92  & 14.39  & 14.22	   \\
11    &	201405.1+412031.2  &     &   201405.08+412031.0\tablenotemark{a}   & 201405.09+412031.0  & \nodata & 1313-0383472  &  & 15.80 & 15.14 & 14.51    \\
12      &	201405.9+411758.9	&  	&	\nodata	  & 201405.99+411759.0  & \nodata & 1312-0382785 &  & \nodata  & \nodata  & \nodata	    \\
15 	&	201408.9+411203.7	&	&   201408.90+411203.7\tablenotemark{b}  &  \nodata  &  \nodata &  1312-0382831 &  & 15.77  & 14.69  & 14.84  \\
17   	&	201410.5+411439.1  &  	&   \nodata   & \nodata  & \nodata  & 1312-0382865  &  &  \nodata  &	 \nodata  &	 \nodata 	 \\
20   	&	201412.2+412056.2  &  	&   \nodata   & \nodata  & \nodata  & 1313-0383588 &  &  \nodata  &	 \nodata  &	 \nodata 	 \\
21 	&	201412.5+411052.0  & 	&   201412.45+411052.0\tablenotemark{b}  &  \nodata  & \nodata  & \nodata &  & 15.34  & 14.71  & 14.45   \\
22 	&	201412.6+411316.0	&	&   201412.60+411316.0  &  \nodata  & \nodata  & 1312-0382903  &  & 13.26  & 12.57  & 12.43 	  \\
23	&	201413.5+411348.9	&	&   201413.46+411349.0  &  \nodata  &  \nodata & 1312-0382911  &  & 14.60  & 13.69  & 13.51 	  \\
24	&	201413.7+411715.5	&	&   201413.77+411716.4  &  201413.79+411716.4  & \nodata  &  1312-0382915 &  & 12.99  & 12.25  & 12.00 	  \\
25   	&	201414.1+412106.9  &  	&   \nodata   & \nodata  & \nodata  & \nodata &  &  \nodata  &	 \nodata  &	 \nodata 	 \\
26	&	201416.3+411620.5	&	&   201416.34+411620.6  &  \nodata  & \nodata  &  1312-0382949 &  & 13.95  & 13.16  & 12.86    \\
28	&	201418.1+411512.7	&	&   201418.06+411512.6  &  \nodata  &  \nodata &  1312-0382981 &  & 13.63  & 12.83  & 12.51   \\
30	&	201419.1+411813.4	&	&   \nodata  &  \nodata  & G78.179+3.564 & \nodata  &  & \nodata  & \nodata  & \nodata 	 \\
31	&	201419.4+411304.9	&	&   201419.47+411303.3  &  \nodata  &  \nodata &  \nodata &  & 11.65  & 11.32  & 11.20  \\
32	&	201419.8+411155.9	&	&   201419.82+411155.8  &  \nodata  &  \nodata & 1311-0384874  &  & 12.73  & 12.18  & 11.99  \\
33	&	201420.0+411415.3	&	&   201419.99+411415.4\tablenotemark{a}  &  \nodata  &  \nodata & \nodata  &  & 16.25  & 14.44  & 13.78 	\\
34 	&	201420.4+412136.9  &  	&   \nodata   & \nodata  & \nodata  & 1313-0383733  &  &  \nodata  &	 \nodata  &	 \nodata 	 \\
36	&	201423.5+410853.3	&	&   201423.44+410853.6\tablenotemark{b}  &  \nodata  & \nodata  & 1311-0384945  &  & 15.83  & 14.77  & 14.49  \\
37	&	201425.0+411332.0	&	&   201424.98+411332.0  &  \nodata  & \nodata  & 1312-0383063  &  & 12.06  & 11.68  & 11.49 	 \\
38	&	201425.2+411809.1	&	&   201425.24+411809.0  &  \nodata  &  \nodata &  1313-0383828 &  & 14.94  & 14.07  & 14.07 	  \\
39	&	201425.3+411517.9	&	&   201425.35+411517.9  &  \nodata  &  \nodata & 1312-0383070  &  & 13.45  & 12.66  & 12.41 	 \\
40    &	201425.6+411305.7	&    &   201425.59+411305.5\tablenotemark{a}   & 201425.60+411305.4  & \nodata & \nodata  &  & 17.90  & 17.38  & 14.72	    \\
41	&	201425.9+411336.5	&	&   201425.86+411336.3\tablenotemark{b}  & \nodata   &  \nodata & \nodata &  & 14.56  & 12.77  & 10.34    \\
42	&	201426.0+411331.7  &	&  \nodata      &  \nodata   & G78.1901+3.6329 &  \nodata &  & \nodata & \nodata & \nodata  \\
43	&	201426.2+411327.9  &	&  \nodata      &  \nodata   & G78.1896+3.6317 &  \nodata &  & \nodata & \nodata & \nodata  \\
45	&	201426.3+411802.7	&	&   201426.35+411802.6  &  \nodata  & \nodata  &  1313-0383849 &  & 13.16  & 12.50  & 12.32   \\
47	&	201426.6+411210.2	&	&   201426.61+411210.1  &  201426.63+411210.2  &  \nodata & 1312-0383087  &  & 9.59   & 9.46 	 & 9.39 	    \\
48	&	201426.7+411300.9	&	&   201426.69+411300.9  & \nodata   &  \nodata & 1312-0383088  &  & 15.20  & 14.27  & 13.9 	   \\
49	&	201426.8+411343.0	&	&   201426.81+411342.9  &  201426.81+411342.8  & \nodata  & 1312-0383089  &  & 13.22  & 12.32	  & 11.98 	    \\
50	&	201427.0+411328.7  &	&   201426.90+411328.8\tablenotemark{a}  &  \nodata   & G78.1911+3.6298 &  \nodata &  & 14.39 &12.97 & 11.85  \\
51	&	201427.1+411516.8	&	&   201427.10+411516.6  &   \nodata & \nodata  & 1312-0383094  &  & 13.66  & 12.92  & 12.71    \\
52	&	201427.4+411533.1	&	&   201427.37+411533.1  &  \nodata  &  \nodata & 1312-0383096  &  & 12.92  & 12.26  & 12.00	  \\
53	&	201427.4+410944.4	&	&   201427.38+410944.0  &  \nodata  & \nodata  & 1311-0385021  &  & 14.75  & 13.78  & 13.44   \\
55	&	201428.0+411303.4	&	&   201427.91+411303.3\tablenotemark{b}  &  \nodata  & \nodata  & \nodata  &  & 16.36  & 14.95  & 14.12    \\
56	&	201428.1+410954.2	&	&   201428.08+410953.9  &  \nodata  & \nodata  & 1311-0385036  &  & 14.89  & 14.09  & 13.73     \\
57	&	201428.2+411317.3	&	&   201428.17+411317.0  &  \nodata  &  \nodata & \nodata  &  & 15.85  & 14.64  & 14.19 	   \\
58	&	201428.4+411351.8	&	&   201428.38+411351.7  &  201428.39+411351.7\tablenotemark{c}  & \nodata  & 1312-0383110  &  & 12.67  & 11.56  & 10.82 	\\
61	&	201429.7+411444.2	&	&   201429.73+411444.1\tablenotemark{b}  &  \nodata  &  \nodata & \nodata  &  & 16.22  & 15.23	  & 15.12    \\
62	&	201429.8+411309.2	&	&   201429.78+411309.2  &  \nodata  & \nodata  & 1312-0383133 &  & 13.78  & 12.97  &  12.64  	   \\
63	&	201429.8+411322.8	&	&   201429.76+411322.9  &  \nodata  & \nodata  & \nodata  &  & 15.35  & 14.29  & 13.87  	   \\
65	&	201430.4+411430.7	&	&   201430.37+411430.5  &  \nodata  &  \nodata & \nodata &  & 14.10  & 13.04  & 12.67 	  \\
66	&	201430.9+411321.9	&	&   201430.86+411322.0  &  \nodata  & \nodata  &  1312-0383158 &  & 14.38  & 13.61  & 13.33 	  \\
67	&	201431.6+411219.5	&	&   201431.62+411219.6\tablenotemark{b}  &  \nodata  & \nodata  & \nodata  &  & 15.94  & 14.87  & 14.52 	 \\
69	&	201432.1+411249.6	&	&   201432.05+411249.8  &  \nodata  & \nodata  & 1312-0383171  &  & 9.97  & 9.76  & 9.66 	  \\
70	&	201432.5+411453.8	&	&   201432.46+411453.7  &  \nodata  & \nodata  & 1312-0383178  &  & 8.11  & 7.87  & 7.84    \\ 
72	&	201433.6+411230.5	&	&   201433.58+411230.6\tablenotemark{a}  &  \nodata  & \nodata  & \nodata &  &14.04  & 12.99  & 12.49 	  \\
75    &	201434.0+411212.9  &  	&   \nodata   & \nodata  & \nodata  & 1312-0383209  &  &  \nodata  &	 \nodata  &  \nodata  \\
76	&	201434.3+411150.3	&	&   201434.33+411150.4  &  \nodata  &  \nodata & 1311-0385154  &  & 14.30  & 13.34  & 12.96    \\
77	&	201434.5+411255.0	&	&   201434.52+411255.0  &  \nodata  & \nodata  & 1312-0383220  &  & 13.47  & 12.62  & 12.29   \\
78	&	201434.6+411059.8	&	&   201434.67+411059.5\tablenotemark{b}  &  \nodata  & \nodata  & 1311-0385158 &  & 16.52  & 14.98  & 14.25 	  \\
79	&	201434.8+411748.6	&	&   201434.82+411748.7  &  \nodata  & \nodata  & 1312-0383226  &  & 12.29  & 11.99  & 11.91   \\
81	&	201436.2+411726.5	&	&   201436.26+411726.5  &  \nodata  & \nodata  & 1312-0383241  &  & 15.43  & 14.53  & 14.33     \\
83	&	201437.1+411521.6	&	&   201437.13+411521.7\tablenotemark{a}  &  \nodata  & \nodata  &  \nodata  &  & 15.66  & 15.68  & 14.60 	  \\
84	&	201437.1+411343.3	&	&   201437.11+411343.0  &  \nodata  & \nodata  & \nodata  &  & 15.35  & 14.38  & 13.97 	  \\
85	&	201437.5+411020.4	&	&   201437.51+411020.4  &  \nodata  &  \nodata & 1311-0385214  &  & 14.70  & 13.91  & 13.66 	   \\
86	&	201437.9+411111.8	&	&   201437.85+411111.7  &  \nodata  & \nodata  & 1311-0385221 &  & 13.46  & 12.35  & 11.62 	  \\
87	&	201438.4+411302.6	&	&   201438.43+411302.4  &  201438.44+411302.5  &  \nodata & 1312-0383281 &  & 14.75  & 13.72  & 13.39 	     \\
88	&	201439.4+411125.5	&	&   201439.41+411125.4  &  \nodata  & \nodata  & \nodata  &  & 15.99  & 14.86  & 14.29    \\
89	&	201439.6+411446.3	&	&   201439.56+411446.5  &  201439.57+411446.5  & \nodata  & 1312-0383298  &  & 13.13  & 12.52  & 12.28 	    \\
90	&	201440.0+411901.5	&	&   201440.05+411901.5  &  \nodata  & \nodata  & 1313-0384062  &  & 14.38  & 13.55  & 13.26 	   \\
91	&	201440.4+411429.7	&	&   201440.43+411429.5  &  201440.43+411429.4  & \nodata  & 1312-0383314  &  & 12.80  & 12.09  & 11.86 	     \\
92	&	201440.6+411627.2	&	&   201440.57+411627.3  &  \nodata  & \nodata  & 1312-0383316  &  & 15.53  & 14.49  & 14.21   \\
93    &	201440.8+410849.6  &  	&   \nodata   & \nodata  & \nodata  & 1311-0385272  &  &  \nodata  &	 \nodata  &	 \nodata 	 \\
94	&	201440.9+411037.0	&	&   201440.89+411036.9  &  201440.91+411036.8  & \nodata  & \nodata  &  & 15.18  & 14.16  & 13.88 	     \\
95	&	201441.0+411409.8	&	&   201441.03+411409.8  &  \nodata  & \nodata  &  \nodata &  & 14.34  & 13.34  & 13.06 	     \\
96    &	201441.2+410443.4  &  	&   \nodata   & \nodata  & \nodata  & 1310-0385723  &  &  \nodata  &	 \nodata  &	 \nodata 	 \\
97	&	201442.5+412031.0	&	&   201442.50+412030.4  &  \nodata  & \nodata  & 1313-0384111  &  & 14.49  & 13.42  & 12.96 	    \\
100	&	201443.4+411340.5	&	&   201443.33+411340.7  &  201443.33+411340.6\tablenotemark{c}  & \nodata  &  1312-0383363 &  & 13.94  & 12.99  & 12.58 	    \\
103	&	201444.5+411259.6	& 	&   201444.46+411259.7  &  201444.48+411259.5\tablenotemark{c}  & \nodata  & 1312-0383380 &  & 15.49  & 14.40  & 13.81 	     \\
104	&	201444.8+410721.8	&	&   201444.76+410721.5  &  201444.77+410721.4  &  \nodata & 1311-0385341 &  & 11.31  & 11.07  & 10.96 	   \\
105	&	201445.3+411539.0	&	&   201445.29+411538.9\tablenotemark{a}  &  \nodata  & \nodata  & 1312-0383399  &  & 16.58  & 15.87  & 16.05 	   \\
106	&	201445.9+411337.5	&	&   201445.87+411337.4  &  \nodata  &  \nodata & 1312-0383405  &  & 13.44  & 12.91  & 12.71 	   \\
107	&	201445.9+410404.9	&	&   201445.95+410404.6  &  \nodata  &  \nodata & 1310-0385788  &  & 13.63  & 12.84  & 12.63 	   \\
108	&	201449.9+411302.0	&	&   201449.91+411302.1  &  \nodata  & \nodata  & 1312-0383475  &  & 12.53  & 11.93  & 11.66	    \\
109  &	201450.3+410547.2	&  	&	\nodata		   & 201450.42+410547.2  & \nodata & 1310-0385853  &  & \nodata  &  \nodata & \nodata	   \\
110	&	201450.3+411517.1	&	&   201450.31+411516.9  & \nodata   & \nodata  & 1312-0383483  &  & 12.52  & 11.79  & 11.49 	   \\
111	&	201450.4+410906.9	&	&   201450.40+410906.9\tablenotemark{b}  &  \nodata  &  \nodata & \nodata  &  & 12.30  & 11.65  & 11.45 	  \\
112	&	201451.4+411948.0	&	&   201451.36+411947.0  & \nodata   &  \nodata & 1313-0384226  &  & 14.46  & 13.56  & 13.34 	  \\
113	&	201452.4+411340.8	&	&   201452.46+411341.1\tablenotemark{a}  &  \nodata  & \nodata  & 1312-0383521  &  & 16.33  & 15.57  & 15.25     \\
114	&	201453.5+411312.8	&	&   201453.48+411312.4  &  \nodata  & \nodata  &  1312-0383542 &  & 14.07  & 12.98  & 12.50 	  \\
115	&	201453.8+411158.6	&	&   201453.91+411158.4  & 201453.92+411158.2   &  \nodata & 1311-0385540 &  & 14.14  & 13.31     \\
116   &	201453.9+411512.1  &  	&   \nodata   & \nodata  & \nodata  & 1312-0383545  &  &  \nodata  &	 \nodata  &	 \nodata 	 \\
117	&	201454.7+411917.5	&	&   201454.58+411916.8  &  \nodata  & \nodata  & 1313-0384278 &  & 14.82  & 13.97  & 13.55   \\
120	&	201457.0+411618.8	&	&   201457.00+411618.2  &  \nodata  & \nodata  & 1312-0383600  &  & 12.96  & 12.53  & 12.33 	   \\
122	&	201457.9+410909.5	&	&   201457.91+410909.5  &  201457.92+410909.6  & \nodata  & 1311-0385617  &  & 14.61  & 13.69  & 13.26 	    \\
126	&	201459.2+411343.8	&	&   201459.21+411345.4  &  \nodata  & \nodata  & 1312-0383642  &  & 10.93  & 10.67  & 10.53 	  \\
127	&	201501.3+411301.2	&	&   201501.35+411300.7  &  \nodata  & \nodata  & 1312-0383672  &  & 14.83  & 13.88  & 13.64 	 \\
129	&	201502.2+411925.5	&	&   201502.17+411925.7  &  \nodata  & \nodata  & 1313-0384391  &  & 14.34  & 13.48  & 13.22	  \\
131	&	201502.7+411556.2	&	&   201502.71+411555.4  &  \nodata  & \nodata  & 1312-0383692  &  & 12.66  & 12.01  & 11.73 	\\
138  &	201504.5+411026.4	&  	&   \nodata   & \nodata  & \nodata  & 1311-0385734  &  &  \nodata  &	 \nodata  &	 \nodata 	 \\
140	&	201505.8+411619.3	&	&   201505.84+411619.1  &  \nodata  & \nodata  & 1312-0383748  &  & 12.74  & 12.24  & 12.06 	   \\
143  &	201508.8+411337.8	&  	&   \nodata   & \nodata  & \nodata  & 1312-0383790  &  &  \nodata  &	 \nodata  &	 \nodata 	\\
144	&	201509.1+410945.0	&	&   201509.03+410944.2  &  \nodata  & \nodata  & 1311-0385819  &  & 14.07  & 13.34  & 13.16    \\
145	&	201510.1+411439.7	&	&   201510.23+411439.6  &  \nodata  & \nodata  & 1312-0383808  &  & 12.20  & 11.90  & 11.86    \\
146	&	201513.0+411336.7	&	&   201513.06+411336.8\tablenotemark{a}  &  \nodata  & \nodata  & 1312-0383851  &  & 11.46  & 11.19  & 11.07     \\
147	&	201513.2+411121.9	&	&   201513.34+411122.6  &  \nodata  & \nodata  & 1311-0385895  &  & 12.17  & 11.43  & 11.06 	   \\
148	&	201514.4+411531.9	&	&   201514.37+411531.8  &  \nodata  &  \nodata & 1312-0383868  &  & 8.16  & 7.81  & 7.74 	   \\
149  &	201515.3+411107.4  	&  	&   \nodata   & \nodata  & \nodata  & 1311-0385921  &  &  \nodata  &	 \nodata  &	 \nodata 	 \\

\enddata

   \tablenotetext{a}{Theses sources have in the 2MASS band one or two filter detections with quality flag U. This means that the source is not detected in this band or it is detected, but not resolved in a consistent fashion with other bands. The 2MASS data for those sources are not used for the cluster analysis.}

   \tablenotetext{b}{The majority of the 2MASS counterparts have a quality flag A (high-quality photometry) in the three bands, except for 9 of them. Photometric quality flag listed for those 9 sources respectively for the J, H and K$_s$ bands: source 15 AAC, source 21 AAB, source 36 AAB, source 41 AAE, source 55 BAA, source 61 BAC, source 67 AAB, source 78 BAA, source 111 AEA.}
   
   \tablenotetext{c}{Same Spitzer counterparts found as in Qiu et al. (2008)}

\end{deluxetable}

\end{document}